\documentclass[prl,aps,twocolumn,preprintnumbers,floatfix, amsmath,amssymb]{revtex4}

\usepackage[dvips]{graphicx}

\usepackage{amssymb,amsfonts,amsmath}

\usepackage[english]{babel}

\usepackage{fancyhdr}
\begin{document}

\title{A graphical theory of competition on spatial resource gradients}

\author{Alexei B. Ryabov}
\affiliation{ICBM, University of Oldenburg, 26111 Oldenburg, Germany}
\author{Bernd Blasius}
\affiliation{ICBM, University of Oldenburg, 26111 Oldenburg, Germany}

\begin{abstract}
Resource competition is a fundamental interaction in natural communities.
However little is known about competition in spatial environments where 
organisms are able to regulate resource distributions.
Here, we analyze the competition of 
two consumers for two resources in a one-dimensional habitat in which the resources are supplied from opposite sides. 
We show that the success of an invading species crucially depends on the 
slope of the resource gradients shaped by the resident.
Our analysis reveals that parameter combinations which lead to coexistence in a uniform environment may favor alternative stable states in a spatial system, and vice versa. 
Furthermore, differences in growth rate, mortality or dispersal abilities allow a consumer to coexist stationarily with - or even outcompete - a competitor with lower resource requirements.   
Applying our theory to a phytoplankton model, we explain shifts in the community structure that are induced by environmental changes.
\end{abstract}

\maketitle

{\noindent





Correspondence email: a.ryabov@icbm.de, blasius@icbm.de

{\bf Ecology Letters (2011) 14: 220--228}

}

\section*{INTRODUCTION}

Competition for limiting resources is one of the most important
species interactions in ecology and has long been considered as a major driver for 
shaping the spatial structure of communities and limiting species richness.
A graphical theory of resource competition was advanced by MacArthur (1972), Le\'{o}n \& Tumpson (1975) and Tilman (1980, 1982) and revolves around the assumption that consumers that reduce limiting resources to the lowest level will exclude all other competitors 
(the $R^*$-rule).
Being confirmed experimentally (Miller et al. 2005, but see also Wilson et al. 2007), this theory provides a fundamental framework for interpreting the relationship between organisms and their 
shared resources in a uniform environment (Grover 1997; Chase \& Leibold 2003).
Extensions of competition theory to spatial settings 
(Chesson 2000a, 2000b; Klausmeier \& Tilman 2002; Amarasekare 2003)
can be grouped into different classes. 
In spatially homogeneous environments, coexistence can be mediated by trade-offs in life history parameters (Levins \& Culver 1971; Levin 1974; Kneitel \& Chase 2004).
In heterogeneous environments regional coexistence can be mediated by spatial segregation 
(Tilman 1982; Gross and Cardinale 2007), whereas local coexistence may be promoted via source-sink effects (Mouquet \& Loreau 2003; Leibold et al. 2004) or positive correlations between dispersal and competitive abilities (Abrams \& Wilson 2004).
However, previous theory has not thoroughly explored resource competition in continuous spatially variable habitats.

In a system with resource flows, spatial heterogeneity can be created by biotic interactions, as local resource consumption can modify resource levels over larger ranges (Huston \& DeAngelis 1994),
possibly establishing resource gradients over the full extent of the habitat.
The analysis of competition in such systems is mathematically challenging (Grover 1997),
as not only the resource availability, but also the size of favorable areas (Skellam 1951), dispersal rates (Abrams \& Wilson 2004) and many other factors become crucial for the survival of a population (Ryabov \& Blasius 2008).
Despite extensive theoretical studies on
unstirred chemostats (Hsu \& Waltman 1993; Smith \& Waltman 1995; Wu et al. 2004), 
persistence and competition in streams (Speirs \& Gurney 2001; Lutscher et al. 2007),
vegetation patterns in water-limited habitats (Klausmeier 1999; von Hardenberg et al. 2001),
well-mixed systems with a light gradient (Huisman \& Weissing 1994; Diehl 2002), and 
non-uniform phytoplankton systems (Dutkiewicz et al. 2009; Yoshiyama et al. 2009; 
Ryabov et al. 2010),
translating competition theory to extended systems, where species are able to shape resource distributions, still remains a great challenge.

In this study we develop a general framework for analyzing the
competition between two consumers for two limiting resources in a spatially continuous habitat in which resource distributions are regulated by 
biotic interactions.
In extension of the $R^*$-rule for uniform systems,
we introduce the notion of an invasion threshold in a spatial system
as the maximal resource requirement for a consumer to invade 
in the presence of a resident species.
In this way, the outcome of competition can be interpreted graphically 
in the resource plane, by comparing the location of the invader's critical resources with respect to the invasion threshold lines.
We derive analytic expressions for the invasibility conditions
and show that they can be related to how 
resources are spatially reduced as a result of resource-use by the resident.

Using this approach we find that environmental heterogeneity increases the likelihood of coexistence. 
Thereby we identify two forms of stationary coexistence in spatially variable habitats.
First, coexistence can be mediated by a trade-off in resource requirements. 
This is characterized by spatial segregation of the two species and
we find that parameter combinations which allow coexistence in a uniform system
favor alternative stable states in a spatially extended system and
{\it vice versa}. 
Second, coexistence can arise from positive correlations of growth rates and resource requirements
(gleaner-opportunist trade-off), characterized by a lack of spatial segregation.

To illustrate these ideas, we investigate 
two phytoplankton species competing for light and a nutrient in a water column. 
We identify two distinct regimes of bistability in the competition outcome 
and show that our approach provides a powerful basis 
for projecting the influence of environmental changes
on the community composition.

\section*{MODELLING FRAMEWORK AND ANALYSIS}
\subsection*{Resource competition in uniform systems \label{sec:Tilman}}

As a background for the following discussion, we briefly review the theory of resource competition in a uniform environment (MacArthur 1972; Le\'{o}n and Tumpson 1975; Tilman 1980, 1982; Grover 1997; Chase \& Leibold 2003).
Consider two consumer species $i=1,2$  growing on two essential resources $N$ and $I$. The dynamics of the population density $P_i$ of each species is determined by the difference between growth $\mu_i(N,I)$ and mortality  $m_i$
\begin{equation}
\dot{P}_i = (\mu_i (N,I) -m_i)\,  P_i \, , 
\label{eq:basic}
\end{equation}
where the growth rate 
in general can be written as
\begin{equation}
\mu_i(N, I) = \mu_{max,i} \, \, \mu\left( \frac{N}{H_{N, i}}, \frac{I}{H_{I, i}} \right)  .
\label{eq:growth_rate}
\end{equation}
Here $\mu_{max,i}$ is the maximal growth rate of species $i$, 
$\mu$ is a monotonically increasing function of both arguments with upper bound $\lim_{N, I \rightarrow \infty} \mu(N, I) = 1$, and
the half-saturation constants 
$H_{N,i}$ and $H_{I,i}$  define the species' resource adaptation.
In particular, this functional form includes 
von Liebig's law of minimum, where the limitation of growth follows the Monod kinetics 
\begin{equation}
\mu_i(N, I) = \mu_{max,i} \, \,  \min \left\{
\frac{N}{N + H_{N, i}} \ , \ \frac{I}{I + H_{I, i}}
  \right\} \ . 
\label{eq:grrate}
\end{equation}

In a uniform environment, levels of system resources can be represented as a point in a 
two-dimensional resource plane (Fig.~1a). In this plane, 
states of balance between growth and mortality,
$\mu_i (N,I) = m_i$, 
determine the zero net growth isoclines (ZNGIs),
which for each species divide the resource plane into areas of positive and negative population net growth.
Assuming eqn~\ref{eq:grrate} for the growth rate, we can calculate the critical  values, $I^*_i =  H_{I, i}\ m_i/(\mu_{max,i} - m_i)$ and $N^*_i = H_{N, i} \ m_i/(\mu_{max,i} - m_i)$, and the ZNGI takes the 
form of two orthogonal lines (dashed lines in Fig.~1).

The resource concentrations in the absence of consumers define the coordinates of a resource supply point $S$. 
A growing population will deplete resources, shifting the system state in the direction of the consumption vector $CV$, 
until the system state point hits the population's ZNGI
at the equilibrium point 
$E= (\widetilde{N}_i, \widetilde{I}_i)$. 
This point defines the resource levels at equilibrium, 
which are shaped by the population,
and sets the conditions for the invasion of other species.

\begin{figure}[tb]
\begin{center}
\centerline{\includegraphics[width=\columnwidth]{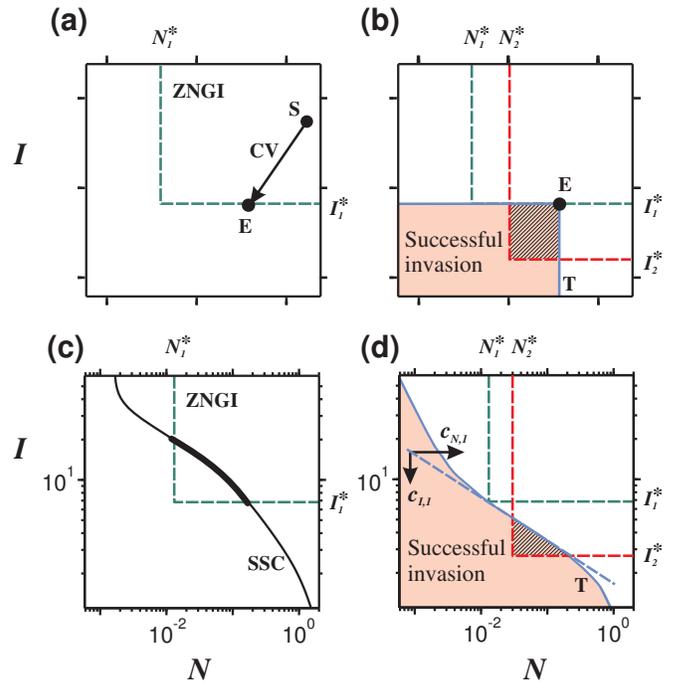}}
\caption{Invasion analysis in a uniform system (top) and a spatially extended  system with opposing resource gradients (bottom).
(a)  Equilibrium configuration of the resident (species 1, green). 
The system state point at equilibrium $E$
is determined by the intersection of the consumption vector $CV$, 
starting at the resource supply point $S$, 
with the zero net growth isocline ZNGI (dashed line, see text).
(b) Critical resource values  of a successful invader 
(species 2, red)  should be located below the invasion threshold $T$ (blue line). 
The intersection of the invader's ZNGI with $T$ is shown as cross-hatching.
(c) In a spatially extended system, 
combinations of resources at different spatial coordinates give rise to a system state curve SSC (solid line, favorable range is marked in bold).
(d) Invasion threshold $T$ (blue solid line) 
and first order approximation (blue dashed line) 
with slope  $\gamma_1=c_{I,1}/c_{N,1}$ in the log-log plot (eqn~\ref{eq:criteria_IN}).
(c) and (d) show results of
numerical simulations in the phytoplankton model, eqns~3--\ref{eq:1splight} 
(see Appendix S5 for model parameters).}
\label{fig:comp_cl}
\end{center}
\end{figure}

Assume that resident species 1 has attained equilibrium.
Then invasion of a second species 2 is possible if the equilibrium resource concentrations 
shaped by the resident 
are sufficient to allow positive growth of the invader.
Thus, in resource space a part of the invader's ZNGI must be located below the equilibrium point $E$ (Fig.~\ref{fig:comp_cl}b). 
For the specific form of eqn~3, this condition is fulfilled for all critical resource values of the invader that are located in the rectangular area defined by 
$I^*_2 < \widetilde{I}_1$ and $N^*_2< \widetilde{N}_1$.
This line defines the invasion threshold $T$ for species 2 (blue line in Fig.~1b).

Combining the invasion analysis for each of the two species yields the well-known 
outcomes of resource competition in a uniform environment (Tilman 1982).
Without a trade-off in the use of the two resources
(i.e. the two ZNGIs do not intersect), the species with the lowest resource requirements wins.
In the presence of a trade-off, competition can lead 
to stable coexistence, competitive exclusion or bistability 
(i.e. alternative stable states
with outcomes depending on initial conditions),
depending on whether or not each species has a greater impact on the resource that 
most limits its own growth (see Fig.~S1).

\begin{figure*}[t!]
\begin{center}
\centerline{\includegraphics[width=0.9\textwidth]{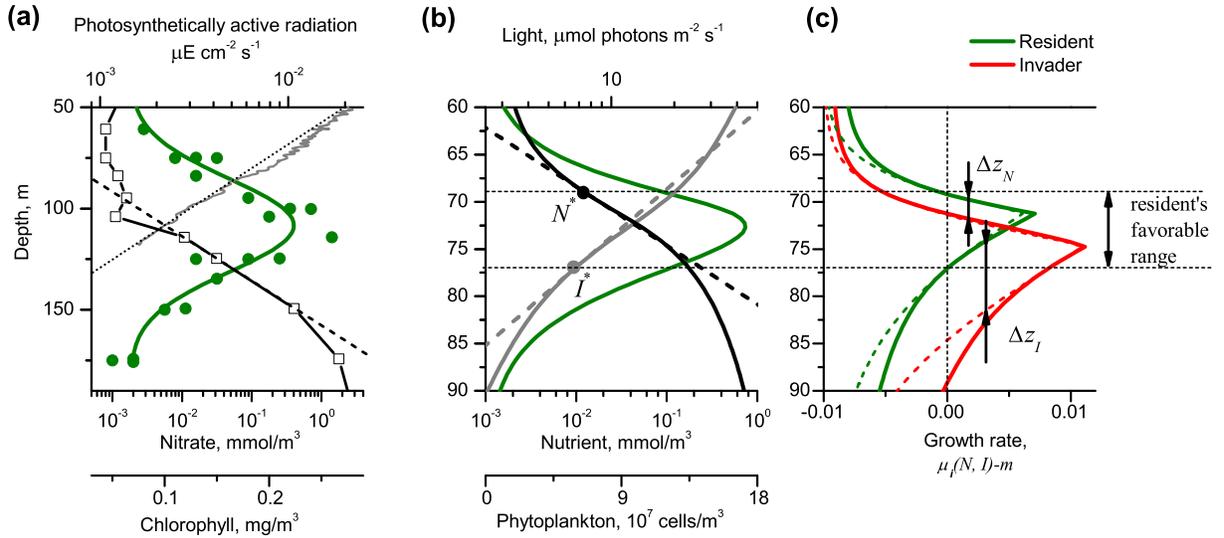}}
\caption{
Typical distributions of light (gray), nutrient (black) and chlorophyll/phytoplankton (green) concentrations  in a water column from (a) field measurements (HOT program, Station ALOHA, cruise 114; see also Karl 2010) and (b) numerical simulation of eqns~\ref{eq:grrate}--\ref{eq:1splight}.
Dashed lines in (a) and (b) indicate exponential fit of the resource distributions.
(c) Net growth rates, $\mu_i(N, I) - m$, of a resident (species 1, green) and an invader (species 2, red), derived from the simulated 
distribution of resources (solid lines) and from their exponential approximation 
(eqn~7, dashed lines). 
Horizontal dashed lines in (b) and (c) mark the favorable range of species 1.
Vertical shifts $\Delta z_N$ and $\Delta z_I$, related to nutrient and light limitation (eqn 10), are also shown.
Model parameters as in Fig.~1.}
\label{fig:1distr}
\end{center}
\end{figure*}

\subsection*{Resource distributions in a spatial system}


The classical approach implicitly assumes the presence of boundaries which confine the system,
between which organisms and resources are uniformly distributed.
In the following we aim to translate these ideas to
an open system where the favorable area of a species
is confined by resource availability, and where organisms and resources are not distributed uniformly.
Consider a one-dimensional environment, where
the population densities $P_i(z, t)$ depend on a spatial coordinate $z$. 
A spatial analog to eqn~1 can be written in the form
\begin{equation}
\frac{\partial P_i}{\partial t} =
  (\mu_i(N, I) - m_i ) \, P_i + D_i \frac{\partial^2 P_i}{\partial z^2} \ ,
\label{eq:basic_space}\\
\end{equation}
where $D_i$ characterizes the diffusive dispersal ability of population $i$. 
Again the growth rate $\mu_i(N, I)$ follows the general form of
eqn~\ref{eq:growth_rate}, but the resource concentrations $N(z,t)$ and $I(z,t)$ are functions of $z$,  so that the two populations are indirectly coupled through 
the spatial profile of their shared resources.

As a case study, we examine
a two-species phytoplankton-nutrient model in an incompletely mixed water column
(Radach \& Maier-Reimer 1975, Jamart et al. 1977, Klausmeier \& Litchman 2001, Huisman et al. 2006, Ryabov et al. 2010).
In the model, eqn~4 describes the density $P_i(z, t)$ of phytoplankton species $i$ at depth $z$ and time $t$,
and is complemented by an equation for the nutrient $N(z, t)$ 
\begin{equation}
\frac{\partial N}{\partial t} =
 - \sum_{i=1}^n \alpha_i \mu_i(N, I) P_i + 
D\frac{\partial^2 N}{\partial z^2}  \label{eq:1spntr} \ ,
\end{equation}
and an equation for the light intensity $I(z, t)$, which describes the absorption of light by water and phytoplankton (Kirk 1994)
\begin{equation}
I(z) = I_{in} \exp \left[ -K_{bg} z - \int^z_0 \sum_{i=1}^n k_i P_i(\xi, t) d \xi \right] \ .
\label{eq:1splight}
\end{equation}
In these equations,
parameter $D$ obtains the role of the turbulent diffusivity,
$\alpha_i$ is the 
nutrient content of a phytoplankton cell,
$I_{in}$ is the incident light intensity, $K_{bg}$ is the water turbidity and $k_i$ is the attenuation coefficient of phytoplankton cells. 
As boundary conditions we assumed impenetrable borders at the surface and at the bottom for the phytoplankton biomass and an impenetrable surface and a constant concentration $N_B$ at the bottom for the nutrient (for model parameters see Table S1).


The model describes a situation in which
the two resources, light and nutrient, are supplied from opposite sides of a spatial habitat. 
This is typical in the ocean where light is supplied from above and many macronutrients from below (Fig.~2a).
As shown in Fig.~2b,
this gives rise to characteristic resource distributions with inverse gradients,
where population growth is maximal at an intermediate position at which both resources are sufficiently available (see Fig.~S6 for further typical model solutions).

Within the favorable range of the population,
the light and nutrient distributions in Figs.~2a and 2b
can be well approximated as straight lines on a logarithmic scale.
This means that resource distributions at equilibrium decay (or grow) exponentially in space 
\begin{equation}
  \widetilde{N}_i(z)  \sim  e^{c_{N,i} \, z}, \ \ 
  \widetilde{I}_i(z)  \sim  e^{- c_{I,i}\,  z } \, .
\label{eq:exp}
\end{equation}
This exponential dependence can be derived analytically in the limit of 
low mortality (see Appendix S1), and it is typical for many other ecosystems 
(see e.g. Fig.~S5).
Eqn~7 implies that a population $i$ in monoculture is ``shading''
resources by a constant percentage
\begin{equation}
  c_{N,i} =  \frac{1}{N_i}\frac{dN_i(z)}{dz}, \quad
  c_{I,i} = - \frac{1}{I_i}\frac{dI_i(z)}{dz} \ .
\label{eq:logDeriv}
\end{equation}
Here, the logarithmic resource gradients $c_{N,i}$ and $c_{I,i}$ 
measure the influence of the population on the resource distribution.

Translated into the resource plane, 
the resource distributions of the spatially extended system can be compactly represented 
as a parametric curve, $(N(z), I(z))$,
(solid line in Fig.~1c).
This `system state curve' (SSC) naturally extends the notion of a system state point E of a homogeneous system to a spatially continuous environment  (Ryabov et al. 2010). 
As shown in Fig.~1c, due to the spatial coupling and source-sink effects the system state curve at equilibrium does not settle at the ZNGI.
Instead, it intersects the ZNGI at two points which mark the boundaries of positive net growth.  This inner part of the system state curve (marked in bold in Fig.~1c) corresponds to the resident's favorable range (horizontally dashed lines in Fig.~2b, c).
Given the exponential dependence (eqn~7), within the favorable range 
the system state curve approaches a straight line 
with slope $\gamma_i = c_{I,i}/c_{N,i}$
in the resource plane with logarithmic axes
(see Fig.~1c in the model or Fig.~S4c for field data).

\subsection*{Invasion analysis in a spatial system}

In analogy to competition theory in a uniform system, we may now ask whether it is possible to predict the outcome of two-species competition from knowledge about equilibrium resource distribution which has been shaped by each species alone in the absence of the other species. 
Assuming that we know the resident's system state curve at equilibrium  $(\widetilde{N}_1(z), \widetilde{I}_1(z))$, what can we say about the 
success of an invader (red color in Fig. 2c)?

To address this problem we 
define the invasion threshold T in the resource plane as the 
set of critical resources of all invaders of small initial density,
which have zero total growth 
in the presence of the resident  (blue solid line in Fig.~1d,
obtained by invasion analysis in numerical simulations with 7000 parameter combinations).
In this formulation, the success of an invader depends on whether or not its ZNGI intersects with the invasion threshold (cross-hatching in Fig.~1).
Thus, invasion analysis can be performed graphically by comparing the location of the invader's critical resources with the invasion threshold.

By comparison of Figs.~1b and 1d, the invasion threshold has a 
different shape in a uniform or a spatially continuous system.
It is difficult to predict the invasion threshold from general principles.
Exact conditions can be expressed in terms of eigenvalues of a corresponding boundary value problem (Hsu \& Waltman 1993; Appendix S3), which unfortunately cannot be solved in general.
In the following we show 
that, with the assumption of exponential resource distributions (eqn~7), it is possible to calculate the invasion threshold for species with similar resource requirements.

\subsection*{Calculation of the invasion threshold}

Consider first the simplest case where resident (species 1) and invader (species 2) 
differ only in their half saturation constants, $H_{N,i}$ and $H_{I,i}$, but
are otherwise identical
($\mu_{max, 1} = \mu_{max, 2}$, $m_1 = m_2$,  $D_1 = D_2$).
As the invader density is assumed to be small, and 
its influence on the resource distributions can thus 
be neglected, the possibility of invasion
depends entirely on the invader's growth rate 
in the equilibrium resource distribution shaped by the resident,
$\mu_2(\widetilde{N}_1(z), \widetilde{I}_1(z)) $.
The mathematical identity 
\begin{equation}
\frac{1}{H_2}  e^{c z} =
\frac{1}{H_1}  e^{c(z - \Delta z)}\ ,\quad \mbox{with}  \   \Delta z = \frac{1}{c} \ln \frac{H_2}{H_1}  \ , 
\label{eq:trick}
\end{equation}
shows that division of an exponential function by different 
(half-saturation)
constants $H_1$ and $H_2$ is equivalent to a shift $\Delta z$ in position along the $z$-axis. 
Therefore, assuming that the growth rate follows eqn~\ref{eq:growth_rate} and the resource distributions are exponential (eqn~\ref{eq:exp}),
it is possible to express the growth rate, $\mu_2(z)$, of the invader 
through that of the resident 
\begin{equation}
\mu_2(\widetilde{N}_1(z), \widetilde{I}_1(z)) = \mu_1(\widetilde{N}_1(z - \Delta z_N), \widetilde{I}_1(z - \Delta z_I)) \ .
\end{equation}
Here the shifts $\Delta z_N = c_{N, 1}^{-1} \ln H_{N, 2}/H_{N, 1}$ and $\Delta z_I=-c_{I, 1}^{-1} \ln H_{I, 2}/H_{I, 1}$ have opposite signs, since the resource gradients are inverse. 

Introducing a new position $z' = z - \Delta z_N$, we find that 
$ \mu_2(z) = \mu_1(\widetilde{N}_1(z'), \widetilde{I}_1(z' + \Delta)) , $
with
$\Delta =  \Delta z_N - \Delta z_I $. 
The change of variables $z \rightarrow z'$ corresponds to a shift along the z-axis 
(see Fig.~2c)
and has no effect on the conditions for survival as long as boundary effects can be neglected.
Thus, the difference in growth between
invader and resident depends only on the value and sign of the single parameter $\Delta$. 
If $\Delta = 0$, 
the distinct resource requirements of the species just result in a parallel translation of the growth rate profile.
Since the resident species at equilibrium has zero total growth, the same holds for the invader, and the population of the invader cannot establish. 
By contrast,  if $\Delta > 0$, we obtain $I(z' + \Delta) < I(z')$ since $I(z)$ decays with $z$. 
This leads to negative net growth of the invader since the function $\mu(N,I)$ 
increases monotonically with its arguments. 
Using the same arguments, $\Delta < 0$ gives rise to positive growth of the invader.

Thus, we can formulate the invasibility criterion in the form
$\Delta < 0$.
The parameter $\Delta$ describes the
invasion threshold and can be interpreted in terms of resource gradients and relative competitive abilities
\begin{equation}
 \Delta \, \,  =
\underbrace{\frac{1}{c_{N, 1}} }_{
\begin{smallmatrix}
  \text{inverse}\\
  \text{log gradient}\\
  \text{in $N$}
\end{smallmatrix}
}
\underbrace{\ln \frac{H_{N, 2}}{H_{N, 1}} }_{
\begin{smallmatrix}
  \text{log of relative }\\
  \text{competitive}\\
  \text{abilities for $N$}
\end{smallmatrix}
}
+ 
\underbrace{\frac{1}{c_{I, 1}} }_{
\begin{smallmatrix}
  \text{inverse}\\
  \text{log gradient}\\
  \text{in $I$}
\end{smallmatrix}
}
\underbrace{\ln \frac{H_{I, 2}}{H_{I, 1}} }_{
\begin{smallmatrix}
  \text{log of relative }\\
  \text{competitive}\\
  \text{abilities for $I$}
\end{smallmatrix}
} \, .
\end{equation}
If for instance, both species have equal resource requirements for $N$,
$H_{N, 2} = H_{N, 1}$,
invasion is only possible, $\Delta < 0$, if the invading species is a better competitor for $I$, i.e.
$\ln (H_{I, 2} / H_{I, 1})<0$.

If the limitation of growth follows von Liebig's law (eqn~\ref{eq:grrate}),
the spatial profiles for light and nutrient limitation 
are shifted independently and  the net shift $\Delta =  \Delta z_N - \Delta z_I$ measures
the difference in the size of the resident and invader's favorable habitats. 
This is depicted in Fig.~2c for a situation with $\Delta z_N < \Delta z_I$, where
the narrowing of the favorable range due to higher $N$ requirements ($N_2^* > N_1^*$) is less than the widening due to better $I$ adaptation ($I_2^*<I_1^*$). Therefore the 
net growth rate of the invader is greater than that of the resident, allowing for successful invasion (see Fig.~2c).
Note that in the example shown in Figs. 1 and 2, the invader, species 2, 
is a good light competitor, $I^*_2 < I^*_1$, but needs higher nutrient concentrations,
$N^*_2 > N^*_1$. 
Therefore it will have maximal production at a deeper depth (Fig. 2c).

\begin{figure*}[t]
\begin{center}
\centerline{
\includegraphics[width=0.9\textwidth]{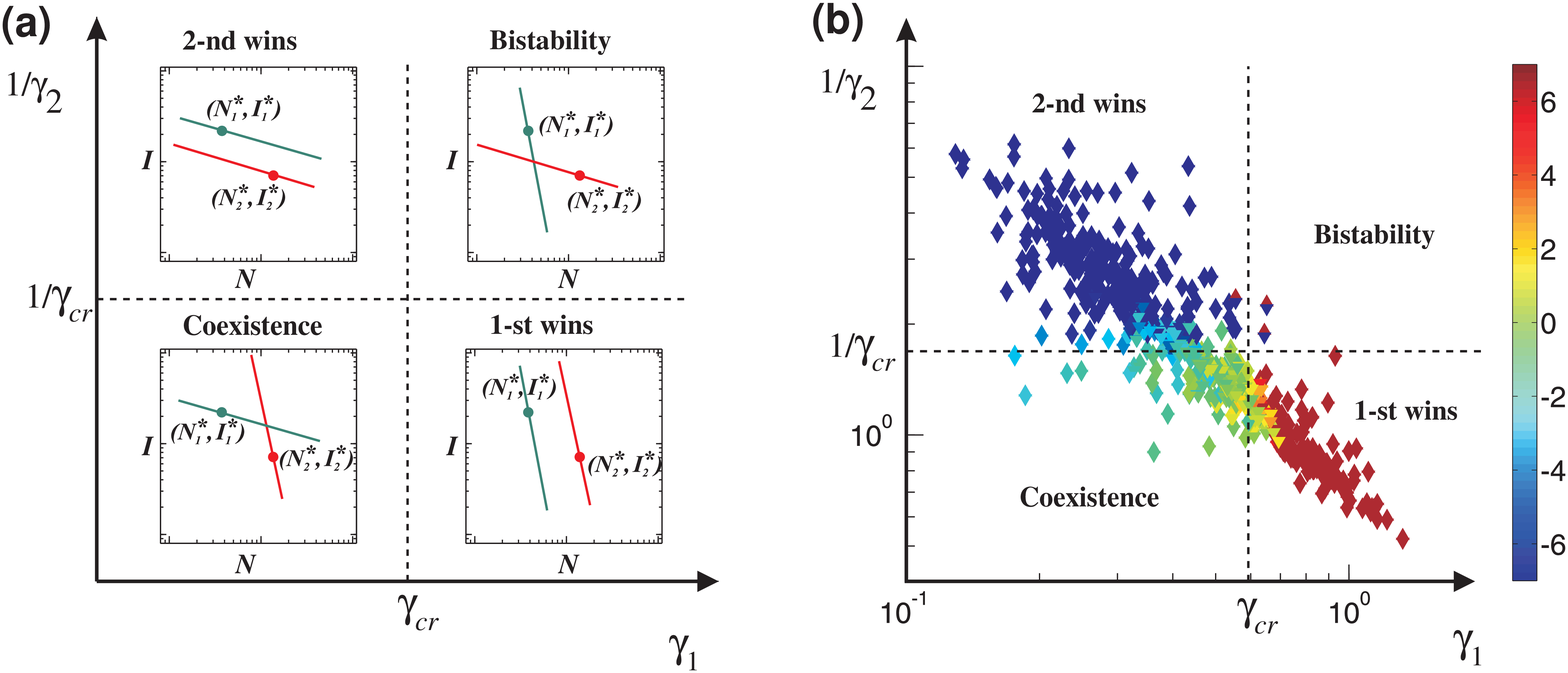}
}
\caption{The four different cases of spatial resource competition 
depending on the 
ratio $\gamma_i$ of logarithmic resource gradients:
($\gamma_1 < \gamma_{cr}$, $1/\gamma_2 < 1/\gamma_{cr}$) coexistence;
($\gamma_1 > \gamma_{cr}$, $1/\gamma_2 > 1/\gamma_{cr}$)
alternative stable states;
($\gamma_1 > \gamma_{cr}$, $1/\gamma_2 < 1/\gamma_{cr}$) species 1 wins;
($\gamma_1 < \gamma_{cr}$, $1/\gamma_2 > 1/\gamma_{cr}$) species 2 wins.
(a) Schematic representation.
Insets sketch invasion threshold lines for resident species 1 (green)
and invader species 2 (red), and critical resource values (dots)
in the resource plane.
(b) Numerical verification in the phytoplankton model, based on
more than 1000 simulations, with parameter values randomly chosen from a wide range 
(critical resources $(N_i^*, \ I_i^*$) and $\gamma_{cr}$ remain constant, see Appendix S5).
The color scale shows the logarithm of relative species abundance $\ln B_1/B_2$ at equilibrium 
(total biomass $B_i = \int_0^{Z_B} P_i(z) \, dz$). 
Upward-pointing triangles indicate the simulation outcome 
when species 1 is the resident and species 2 the invader,
downward-pointing triangles show the opposite case; their color is different only if the
competition outcome is bistable.}
\label{fig:comp}
\end{center}
\end{figure*}

Since the maximal growth and mortality rates are assumed to be equal for both species, the ratio of half-saturation constants equals the ratio of critical resource values (e.g., $H_{I, 2}/H_{I, 1} = I^*_2/I^*_1$), and the invasibility criterion reads
\begin{equation}
\Delta = \frac{1}{c_{N, 1}} \ln \frac{N^*_2}{N^*_1} + \frac{1}{c_{I, 1}} \ln \frac{I^*_2}{I^*_1} < 0 \ .
\label{eq:criteria_IN}
\end{equation}
This expression has a straightforward geometrical interpretation: 
a species can invade the spatial habitat if its critical resources $(N^*_2, \ I^*_2)$ are located below a straight line,
$\ln I - \ln I^*_1 = - \gamma_1 (\ln N - \ln N^*_1)$,
with a slope of absolute value $\gamma_1=c_{I, 1}/c_{N, 1}$ 
passing through the point $(N^*_1, \ I^*_1)$ in the double-logarithmic resource plane 
(blue dashed line in Fig.~\ref{fig:comp_cl}d). 
Thus, we have derived a first-order approximation for the invasion threshold T.

\subsection*{Competition outcome}

In a similar way it is possible to analyze the invasion potential of species 1 in a system with resident species 2.
Assume without loss of generality that  $N^*_2 > N^*_1$ and $I^*_2 < I^*_1$,
and denote by $\gamma_{cr}$ the slope (taken with opposite sign) of a straight
line passing through the two critical resource points, 
$\gamma_{cr}=-(\ln I^*_2/I^*_1)/(\ln N^*_2/N^*_1)$.
Then, combining eqn \ref{eq:criteria_IN} and its counterpart for species 2,
we obtain four different outcomes of spatial resource competition (Fig.~3a).
If the critical resource values for each species are located below the invasion threshold of its competitor, $\gamma_1 < \gamma_{cr}$ and $1/\gamma_2 < 1/\gamma_{cr}$, both species can invade the monoculture of the other,
leading to coexistence.
In the opposite case, if the critical resource values for every species lie above the invasion threshold of its competitor, $\gamma_1 > \gamma_{cr}$ and $1/\gamma_2 > 1/\gamma_{cr}$, 
neither of the two species can invade, leading to bistability.
Finally, one species may be a superior competitor if only this species can grow in the presence of its competitor.

This analytic conclusion is confirmed by numerical simulations
of the phytoplankton model eqns 3-6
in a large parameter range (Fig.~3b).
The figure demonstrates that in spite of the high complexity and nonlinearity of the 
model,
the relation between the slope of the invasion threshold lines, $\gamma_{i}$, and 
the location of the critical resource values
gives a good prediction of the competition outcome.

\subsection*{Differences in growth, mortality and dispersal}

\begin{figure}[tb]
\begin{center}
\centerline{
\includegraphics[width=\columnwidth]{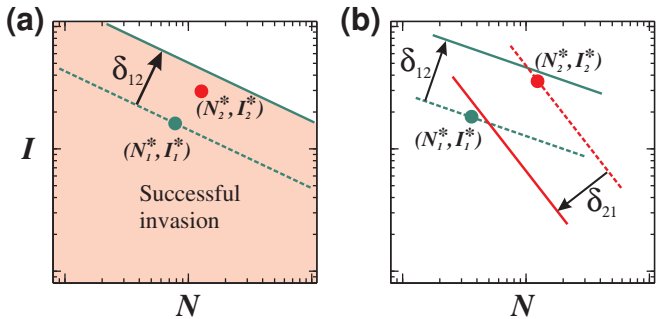}
}
\caption{(a) Interspecific differences in $\mu_{max}$, $m$, or $D$ yield a shift $\delta_{12}$ of the invasion threshold in the normal direction. 
Compared with Fig.~1d, the invader (species 2, red) can establish, even though it has higher resource requirements than the resident (species 1, green). 
(b) Coexistence due to a 
gleaner-opportunist 
trade-off, where both invasion thresholds are shifted in opposite directions.
Species 1 (gleaner) with lower resource requirements can coexist at equilibrium with species 2 (opportunist), which has higher resource requirements but higher $\mu_{max}$, or lower $m$ or $D$.
}
\label{fig:compglopp}
\end{center}
\end{figure}

In general, the locations of the invasion thresholds may depend on trait differences between the resident and invader, such as maximal growth rate, mortality and dispersal rates.  
As shown in Appendix S3, the generalized form of the invasibility criterion reads
\begin{equation}
\frac{1}{c_{N, 1}} \ln \frac{N^*_2}{N^*_1} + \frac{1}{c_{I, 1}} \ln \frac{I^*_2}{I^*_1}  < \Delta_{12}
\label{eq:crithsgr}
\end{equation}
where 
$\Delta_{12}$ 
is a complicated function of the trait values and resource gradients, but 
does not depend on the critical resource levels.
If all trait differences vanish, we obtain $\Delta_{12} =0$
and recover the previous result, eqn~12. 
The geometrical representation of the new term $\Delta_{12}$ is a shift of the invasion threshold line toward 
larger ($\Delta_{12}>0$) or smaller ($\Delta_{12}<0$) 
resource requirements in the normal direction by 
$\delta_{12} = \Delta_{12} c_{N, 1} c_{I, 1}/\sqrt{c^2_{N, 1} + c^2_{I, 1}}$. 
Thus, the slope of the invasion threshold lines remain unchanged, but their location is shifted.
Fig.~4a shows an example with a positive shift of the invasion threshold,
where species 2 can invade the system despite the fact that it has higher requirements for both resources than the resident species 1.

For an illustration of how specific trait values can change the sign of $\delta_{12}$,
consider the particular case that species 2 has higher values of both
growth rate, $\mu_{max,2} = \beta \mu_{max,1}$, 
and mortality, $m_2 = \beta m_1$,  with $\beta>1$.
Since the critical resource values 
$N^*_2 > N^*_1$ and $I^*_2> I^*_1$ are independent of $\beta$, this scaling would not affect the competition outcome in a uniform environment and 
species 2 would not be able to invade the system.
However, in a heterogeneous system the invasion threshold (green line in Fig.~4b) is shifted by
$\delta_{12} > 0$ (see Appendix S3). 
Similar analysis, using rescaling by a factor of $1/\beta$, shows that the threshold for invasion of species 1 in a monoculture of species 2
is shifted in the opposite direction (red line in Fig.~4b). 
This mechanism can lead to stable coexistence of a species 
having high growth rate and resource requirements 
with another species having low growth rate and resource requirements 
(gleaner-opportunist trade-off).

\begin{figure}[t!]
\begin{center}
\centerline{\includegraphics[width=\columnwidth]{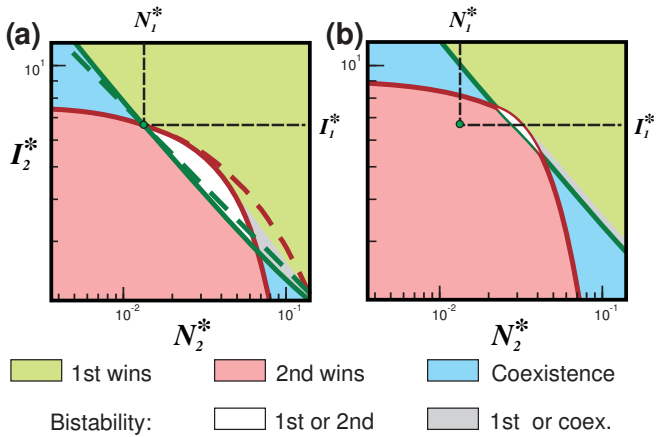}}
\caption{
Competition analysis in the phytoplankton model. 
Maximal growth rates and mortalities are
(a) identical, $\beta = 1$, and (b) different, $\beta=2$. 
The competition outcome (in color coding) of species 1 
(critical resource values indicated as circle) is shown in dependence of the critical resource values of species 2.
Solid lines show invasion thresholds obtained from numerical simulation of eqns \ref{eq:grrate}--\ref{eq:1splight};
dashed lines in (a) show their analytical approximations (Appendix S2). 
Parameter combinations $(N_2^*,I_2^*)$ in the area below the solid green line allow invasion of species 2 in the presence of species 1. In contrast, species 1 can invade in the presence of species 2 with critical resource values taken 
from the area above the solid red line. 
The intersections of these lines define the ranges of coexistence, alternative states, and competitive exclusion. 
}
\label{fig:coexhab}
\end{center}
\end{figure}

\subsection*{Application to a phytoplankton model \label{sec:phyto}}

Applying this approach to a phytoplankton community,
the slopes $\gamma_i$ of the invasion lines 
can be obtained 
by numerical simulations (Fig.~S2), analytic estimations (Appendix~S1), or
in field enclosure experiments (J{\"a}ger et al. 2008).
Tracing the dependence of $\gamma_i$ as a function of system parameters allows us to 
project shifts in the species composition. 
Consider, for instance, the effects of increasing the 
nutrient resource concentration at the bottom of the water column, $N_B$.
More nutrients at the bottom give rise to a larger biomass of the resident species, which in turn 
yields a steeper nutrient decay  within the production layer,
i.e. a reduction of $\gamma_i$ (see Fig.~S2c).
In the resource plane this effectively leads to a counter-clockwise rotation of the invasion threshold line, and 
graphical analysis reveals that by increasing $N_B$ the community composition shifts from dominance of the best nutrient competitor,
through a regime of coexistence,  
to the prevalence of the best light competitor (Fig.~S3). 
This result might explain observed negative correlations of the abundance of high 
light-adapted species with nutrient concentrations (Johnson et al. 2006).

Similarly we can study the influence of consumption rates.
In a well-mixed water column the slope of consumption vectors is given by 
$\Gamma_i = (k_i\, z_B) / \alpha_i$ 
(Huisman \& Weissing, 1994; Diehl 2002).
Thus, the relative resource consumption scales with the ratio of
the light attenuation coefficient to the cell nutrient content, $k_i / \alpha_i$.  
In our model, according to Fig.~\ref{fig:comp}, 
two species coexist if $\gamma_1 < \gamma_{cr} < \gamma_2$.
Since the ratio of resource gradients 
$\gamma_i$ also grows with $k_i / \alpha_i$ (see Fig.~S2a and eqn~S6),
we observe coexistence 
if the best nutrient competitor (here species 1) has a small $k_1 / \alpha_1$ ratio and species 2 a large  $k_2 / \alpha_2$ ratio. 
The reverse
situation (large $k_1 / \alpha_1$ and small $k_2 / \alpha_2$) leads to bistability (Fig.~\ref{fig:comp}a). 
Thus, we obtain a rule which is diametrically opposed
to that for uniform systems:
coexistence arises if each species reduces its least limiting resource more strongly in relation to the other, 
whereas stronger reduction
of the most limiting resource favors bistability (compare Figs.~\ref{fig:comp} and S1).

Fig.~\ref{fig:coexhab}a shows the competition outcome in dependence of the critical resources of species~2 for the case that $k_1 / \alpha_1 > k_2 / \alpha_2$, i.e. species 1 shades relatively more light than species~2. 
In accordance with our rule, if species 2, which has a smaller light attenuation coefficient, 
is more strongly limited by light ($I_2^* > I_1^*$), it may coexist with species 1 (upper blue area), 
whereas stronger nutrient limitation of species 2 
($N_2^* > N_1^*$)
can lead to bistability (white area). However, if the difference in nutrient limitation is 
sufficiently large, then the two species can again coexist (lower blue area, Appendix S2). 
The model further exhibits a regime of bistability between coexistence and single-species dominance (gray area), where the invasion of species 1 does not exclude species 2, but species 2 cannot invade if species 1 has already established 
(see also the vertical profiles in Fig.~S6). 
If the two species differ in their maximal growth and mortality rates, $\beta > 1$
(Fig.~\ref{fig:coexhab}b),
all bifurcation lines are shifted towards larger critical resource values.
Then, if species 2 is parameterized to have higher resource requirements (i.e., critial resource levels located above the black dashed lines in 
Fig.~\ref{fig:coexhab}b), it can have all possible competition outcomes, even without a trade-off in resource requirements.
The corresponding spatial profiles in Fig.~S6 reveal two fundamentally different types of coexistence in a spatial system, mediated either by a trade-off in resource requirements
(characterized by spatial segregation of the two species)
or by a gleaner-opportunist trade-off (lack of spatial segregation).

\begin{table*}
\centering
\begin{tabular}{l|l|l}
\hline \hline
 & Uniform system   &  Spatial system
with resource gradient
        \\ \hline
Invasion threshold  \qquad &  simple form defined by resource values,   
&  curved shape, defined by gradients and resource availability, \\[-0.5ex]
   \ \    \qquad & 
   on the ZNGI                
    & location depends on differences in $\mu_{max}$, $m$, or $D$ \\
\hline
Coexistence  with    & each species mostly consumes 
     & each species mostly consumes its least limiting resource or\\[-1.5ex]   
resource trade-off   & its most limiting resource       &   essential difference in  resource requirements\\[-0.5ex]
        & & $\to$ spatial profiles with separation\\
\hline
Coexistence without  & not possible              &  species with higher resource requirements (opportunist) \\[-1.5ex]  
resource trade-off &   &  has larger $\mu_{max}$, or smaller $m$ or $D$\\[-0.5ex]
        & & $\to$ spatial profiles without separation\\
\hline
Bistability  & 
equally likely as coexistence, & 
less likely than coexistence, \\[-0.5ex]
& each species mostly consumes  & each species mostly consumes its most limiting resource, \\[-1.5ex] 
 & its least limiting resource       & possible also without resource trade-off\\ 
 \hline
\end{tabular}
\caption{
Comparison of competition outcome in 
uniform and spatially extended systems
with inverse resource gradients}
\label{tab:hethomdiff}
\end{table*}

\section*{DISCUSSION}

The extension of resource competition theory to
spatially variable environments 
remains a major challenge for community ecology.
Usual approaches to include environmental heterogeneity consider a range of resource supply points, assuming that the system evolves independently for each supply point (Tilman 1982),
while other studies have concentrated on patch occupancy models (Levins 1979) or metacommunity models of discrete patches that are connected by dispersal (Levin 1974; Mouquet \& Loreau 2003; Abrams \& Wilson 2004; Gross \& Cardinale 2007).
However, links between local patches might not only affect the dispersal of species, but also the spatial flow of matter and resources, leading to meta-ecosystem dynamics (Loreau et al. 2003) 
which require a distinct theoretical approach 
(Huston \& DeAngelis 1994, Smith \& Waltman 1995, Wu et al. 2004).

Our study takes a significant step in this direction by exploring 
competition in a spatially continuous environment where species 
interact indirectly by modifying resource availability beyond their local neighborhood.
We study the critical resource requirements of a successful invader in such a system and find that 
the invasion threshold can be approximated by a straight line on a double-logarithmic scale,
with a slope that is determined by the ratio of logarithmic resource gradients.
Thus, the likelihood of invasion in a two-species community can be understood from the analysis of equilibrium resource distributions in a monoculture of the resident.

Using this approach, 
we clarify the bases of coexistence in spatially variable habitats for a model that has widespread and important applications (see Table~1).
Thereby, we synthesize two previous sets of results concerning coexistence:
i) uniform habitat theory, showing that coexistence can arise from trade-offs in the use of two resources
(Tilman 1980, 1982), and
ii) spatially variable theory showing that coexistence can arise from a 
growth-based trade-off 
(Hsu \& Waltman 1993; Smith \& Waltman 1995; Wu et al. 2004).
We show that in a spatially variable habitat with two resources, both forms of coexistence can occur.

Assuming that two species trade off in their resource adaptation, in a uniform system both species can coexist if each species mostly consumes its most limiting resource (Le\'{o}n \& Tumpson 1975).
In contrast, in a spatial environment 
the species coexist if they each, in monoculture, create a resource distribution with a relatively 
smaller gradient ($c_I$ or $c_N$) 
of their most limiting resource. 
In other words, invasion is possible if the resident species does not ``shade''  its most limiting resource too much. 
As a consequence, parameter combinations which allow coexistence in a uniform system can lead to alternative stable states in a spatially extended system and {\it vice versa}.

These striking differences between resource competition in uniform and spatially variable systems can be explained by the observation that in 
a uniform environment, resources are equally available everywhere, 
whereas in a spatially extended system 
consumers compete for locally available resources. 
Thus in a uniform system, a species can 
suppress invasions
by reducing those resources for which it is the best competitor.
In a spatial setting the same strategy may not work 
because a local reduction of vital resources does not exclude invasions 
elsewhere in the habitat. 
Consider for instance the situation shown in Fig.~2c. Here 
the resident, species 1, is the best 
nutrient competitor,
so that in a uniform environment invasions are prevented if species 1 reduces nutrients more strongly than light (i.e. a small value of $k_1/\alpha_1$, or $\Gamma_1$, in Fig.~S1). In a spatial situation, however, this strategy (small value of $k_1/\alpha_1$)
will promote invasion of species 2 (see Fig.~3), because species 1 can only shade nutrients vertically above its favorable range,
whereas species 2 invades at a greater depth. Therefore, to prevent invasions, species 1 should reduce light as much as possible (large value of $k_1/\alpha_1$) to deteriorate growth conditions in the deep layers.

Applying our theory to a spatial phytoplankton model, we identify two distinct regions of bistability, of either alternative states for each species or bistability between coexistence and a monoculture.
Bistability has been theoretically described in phytoplankton communities 
with differently mixed layers
(Yoshiyama et al. 2009, Ryabov et al. 2010). 
Here we identify bistability in a system with uniform diffusivity, 
however, unlike in well-mixed systems,
the range of bistability is smaller than that of coexistence (see Fig.~5).

Finally, in a spatial system the slope of the invasion threshold
changes with the biomass of the resident (see Appendix~S1), because 
species with higher density have a stronger influence on resource distributions.
In this way, competitive interactions are intricately linked to productivity, which
potentially can give rise to novel causal relations between productivity and diversity (Gross \& Cardinale, 2007; Hillebrand \& Matthiessen, 2009).

\subsection*{Extensions of the theory}

An essential part of our analysis is based on the assumption of 
exponential decay of resources within the favorable range of the resident (eqn~7).
This assumption makes spatial invasion analysis analytically tractable, as it 
allows us to express the growth rate of an invader by that of a resident species, 
thereby circumventing the complicated problem of actually having to calculate spatial density profiles.
Exponential resource distributions correlate with field data 
and can be observed in aquatic systems 
(Figs.~2a and S4; Kirk 1994; Karl \& Letelier 2008), but also appear in other systems, such as marine sediments (Fig.~S5).

With increasing distance from the favorable range one should expect deviations from exponential profiles (Fig.~2b).
Invasion thresholds then become curved (Fig.~1d) and our analytic theory 
only provides a first-order approximation for competition between biologically 
closely related species
(but note that our general approach of investigating invasion thresholds does not have this limitation).  
In this sense our theory complements the approach developed by Yoshiyama et al. (2009) for competition between sufficiently different species.

For simplicity we assumed that the favorable ranges of both species are far
from the system borders, so that boundary effects on the species survival are negligible.
A general approach should include both forms of the invasion threshold, presented in Fig.~\ref{fig:comp_cl}b and \ref{fig:comp_cl}d, as limit cases, 
where the range of positive growth is confined either by 
the boundaries of a well mixed system or by the resource gradients.   
Our numerical simulations in the phytoplankton model (not shown)
demonstrate that alternative stable states replace regimes of coexistence or {\em vice versa}, as the favorable range moves from the interior of the water column to the surface.

The analysis can be extended to take other system parameters into account.
For example, in Appendix S4 we show how sinking or floating can effectively be interpreted as a change of mortality rates, thus also influencing the position of invasion thresholds.
To simplify settings we considered competition between two species under equilibrium conditions.
The approach could be generalized to include temporal changes in habitat conditions.
Resource fluctuations shift SSCs together with invasion thresholds and can provide different time niches for $r$ and $K$ strategists (Grover 1990, 1991). 
A further interesting perspective would be to consider additional species.
The success of a third competitor can be analyzed by including another invasion threshold 
in the resource plane. 
However, analysis of three-species competition is more intricate because it involves the study of
invasion thresholds for three possible pairs of resident species.
In preliminary investigations we have found that the combination of gleaner-opportunist and resource limitation trade-offs may promote stationary coexistence of three species on two resources.

The method developed in this manuscript is independent of the nature of the abiotic resources involved.
Situations with spatial profiles of two inverse chemical resource gradients arise naturally 
whenever resources are provided from two opposite sides, such as interfaces between different environments (D'Hondt et al. 2004).
Moreover, our theory can easily be extended to include
spatial gradients of other vital factors, independently of whether or not they are actually influenced or consumed by the population.
In fact, such a situation is obtained as a special case of our phytoplankton model by setting the light attenuation coefficient $k_i=0$, so that light intensity 
decays exponentially with depth in the water column,  
independently of the biomass distributions. 
Similar spatial variation of growth factors or stress agents is abundant along
environmental gradients and in transition zones between different habitat types.
Thus, the general approach that we develop should apply to 
a wide spectrum of systems, ranging from flow reactors, stream ecosystems and transport-limited vegetation systems to marine sediments and many others.

\section*{Acknowledgments}
We are very grateful to H.J. Brumsack, S. Diehl, C. Dormann, W. Ebenh{\"o}h, B. Engelen, H. Hillebrand, J. Huisman, and J. Meyerholt for discussion of our results.

\bibliographystyle{plain}

\end{document}


\title{SUPPORTING INFORMATION\\}

\maketitle

\section*{S1. Calculation of the logarithmic resource gradients \label{sec:Constants}}

In this section we provide analytic estimates for the light and nutrient gradients 
resulting from a single species population
in the phytoplankton model, eqns 3--6.
We calculate resource distributions 
in the production layer, defined by the critical depths $z_N^*$ and $z_I^*$ 
where either the nutrient or the light intensity reaches a critical value,
$N(z_N^*) = N^*$ and $I(z_I^*) = I^*$ (see dashed lines in Figs.~2b and c).
We further assume small mortality $m \ll \mu_{max}$,
so that the biomass can diffuse from the favorable layer without essential losses and the phytoplankton density has only small variation within the favorable layer,
i.e. $P(z) \approx P_0$ for $z_N^*\leq z \leq z_I^*$. 

From eqn 6 within the favorable range the logarithmic gradient of the light intensity equals
\begin{equation}
c_I = -\frac{d \ln I(z)}{dz} = K_{bg} + k P_0 ,
\label{eq:ci}
\end{equation}
and the light distribution can be written
\begin{equation}
I(z) = I^* e^{ c_I \left( z_I^* - z \right)} \ .
\label{eq:IDecay}
\end{equation}
Regarding the nutrient profile we assume that for small mortality the critical nutrient concentrations are small,
$N^* \ll H_N$, so that
the growth rate close to the critical point $N\approx N^*$ can be linearized 
$
\mu_N(N) \approx \mu_{max} \frac{N}{H_N} \, . 
$
Substituting this expression into eqn 5 we obtain at equilibrium
$$
-\alpha \mu_{max} \frac{N}{H_N} P_0 + D \frac{d N}{d z^2} = 0 \ .
$$
A solution to this equation, that is monotonically increasing with depth, is given by
\begin{equation}
N(z) =  N^* e^{c_N (z - z_N^*)}  \ ,
\label{eq:NDecay}
\end{equation}
with the logarithmic gradient
\begin{equation}
c_N = \frac{d \ln N(z)}{dz} = \sqrt{\frac{\alpha \mu_{max} P_0}{D H_N}} \ .
\label{eq:cn}
\end{equation}

To complete our calculation the phytoplankton density in the production layer is estimated as
$ P_0 \sim \frac{B}{w}$
where $B$ is the total biomass
and the width of the production layer $w = |z_N^* - z_I^*|$ only weakly depends on the model parameters 
(Beckmann and Hense 2007).
If the biomass maximum is located far from the surface, the value of $B$ 
can be estimated in two limiting cases (Klausmeier and Litchman 2001)

\begin{eqnarray}
B = \left\{
\begin{array}{ll}
\frac{\displaystyle \ln (I_{in}/I^*)}{\displaystyle k}
  & \mbox{if }P_0 k \gg K_{bg} \\[8pt]
\frac{D (N_B - N^*)}{\alpha m Z_M} 
  &  \mbox{if } P_0 k \ll K_{bg} \\[2pt]
\end{array} 
\right.
\label{eq:biomass} 
\end{eqnarray}
where  $Z_M =  Z_B - \ln (I_{in}/I^*) / K_{bg}$.

These equations can be further simplified because typically $N_B \gg N^*$. 
Substituting eqns \ref{eq:biomass} into eqns \ref{eq:ci} and \ref{eq:cn},  we obtain the ratio of resource gradients $\gamma = c_I / c_N$ (see also Fig.~\ref{fig:slope})
\begin{eqnarray}
\gamma = \left\{
\begin{array}{ll}
\frac{ \displaystyle K_{bg} +   \frac{ \ln (I_{in}/I^*)}{w}}
{\displaystyle \sqrt{\frac{\alpha}{k} \frac{\mu_{max}}{D H_N}  \frac{ \ln (I_{in}/I^*)}{w} }} 
	 & \quad \mbox{if } P_0 k \gg K_{bg} \\[2pt]
      \mbox{ } & \mbox{} \\
\frac{\displaystyle K_{bg} +  \frac{k}{\alpha}  
\frac{D N_B}{w m  Z_M}}
{\displaystyle \sqrt{\frac{\mu_{max}}{H_N} 
  \frac{ N_B} {w m Z_M} } } 
&  \quad \mbox{if }  P_0 k \ll K_{bg} . \\[2pt]
\end{array} 
\right.
\label{eq:gamma_kbg}
\end{eqnarray}

\medskip

\section*{S2. Bifurcation lines in Fig. 5}

In this section we derive analytic estimates for combinations of critical resource values that allow mutual invasibility of a fixed  ``reference'' species 1 
with a variable ``test'' species 2 (bifurcation lines in Fig.~5a main text).
Both species differ in the values of consumption rates, $k_i$ and $\alpha_i$,
have identical maximal growth rate $\mu_{max}$ and mortality $m$,
but while the critical resource values $N_1^*$ and $I_1^*$ of the 
reference
species are fixed, 
$N_2^*$ and $I_2^*$ are taken from a wide range.

The critical resource values of all test species that can invade in the presence of the reference species are determined by the invasibility criterion, eqn 12
\begin{equation}
 \frac{1}{c_{N, 1}} \ln \frac{N^*_2}{N^*_1} + \frac{1}{c_{I, 1}} \ln \frac{I^*_2}{I^*_1} < 0 \ .
\label{eq:criteriaapp}
\end{equation}
The border of this region is shown as a green dashed line in Fig.~5a.
Similar, we obtain the critical resource values of the test species, which allow the invasion of the reference species when the test species has reached equilibrium 
\begin{equation}
 \frac{1}{c_{N, 2}} \ln \frac{N^*_1}{N^*_2} + \frac{1}{c_{I, 2}} \ln \frac{I^*_1}{I^*_2} < 0 \ .
\label{eq:criteriaappinv}
\end{equation}
In this case however, to plot the bifurcation line, we need to consider the dependence of  $c_{N, 2}$ and $c_{I, 2}$ on the critical resource values, $N_2^*$ and $I_2^*$, of the test species (see. eqns \ref{eq:ci}, \ref{eq:cn} and \ref{eq:biomass}).
For these aims we pick first one special ``hybrid'' test species, that has 
the same critical resource values $N^*_1$, $I^*_1$ as the reference species,
but consumption rates, $\alpha_2$ and $k_2$. 
We denote the logarithmic resources gradients of this species as $c'_{N, 2}$ and $c'_{I, 2}$. 

To proceed we assume that the critical resource values are small:
assuming $N^*_i \ll N_B$ we approximate $N_B - N^*_2 \approx N_B$, and assuming 
$I^*_i \ll I_{in}$ we obtain 
$$
\ln \frac{I_{in}}{I^*_2} \approx \ln \frac{I_{in}}{I^*_1} - O\left(  \frac{I^*_2 - I^*_1}{I_{in}}\right) \ .
$$
Thus, in both limits,
$K_{bg} \gg P_0 k$ and $K_{bg} \ll P_0 k$, in eqn \ref{eq:biomass} the total biomass $B$, and 
from eqn \ref{eq:ci} also the value of $c_{I, 2}$, 
is independent of the critical resource values $N^*_2$, $I^*_2$ of the test species.
This means that
the logarithmic light gradients are identical for all test species
$$
c_{I,2} = c'_{I,2} \ .
$$

In a similar way we estimate the logarithmic nutrient gradients, but from eqn \ref{eq:cn} 
the value of $c_{N,2}$ 
depends on the half-saturation constant of the test species, $c_{N,2} \sim \sqrt{1/H_{N, 2}}$. 
Then the logarithmic nutrient gradient of an arbitrary test species can be expressed in terms of the logarithmic nutrient gradient of the ``hybrid'' species
$$
c_{N,2} = c'_{N,2} \ \sqrt{\frac{H_{N, 1}}{H_{N, 2}}} ,
$$
and using the identity $H_{N,1}/H_{N,2}=N^*_1/N^*_2$
we obtain 
$$
c_{N,2} = c'_{N,2} \ \sqrt{\frac{N^*_1}{N^*_2}} .
$$

Therefore, inequality \ref{eq:criteriaappinv} takes the form
\begin{equation}
 \frac{1}{c'_{N, 2}} \sqrt{\frac{N^*_2}{N^*_1}} \ln \frac{N^*_2}{N^*_1} + \frac{1}{c'_{I, 2}} \ln \frac{I^*_2}{I^*_1} > 0 \ ,
\label{eq:criteriaappinv2}
\end{equation}
where the parameters $c'_{N, 2}$ and $c'_{I, 2}$ are calculated for the ``hybrid'' species having the critical resource values, $N^*_1$ and $I^*_1$, and the consumption rates, $k_2$ and $\alpha_2$. 
Note that this expression holds in both limits when $K_{bg} \gg P_0 k$ and when $K_{bg} \ll P_0 k$,  implying that it might provide a good approximation in general. The border of this region is shown as red dashed line in Fig.~5a.

\section*{S3. Invasion thresholds and principal eigenvalues \label{sec:ev}}

The dynamics of a spatially extended population $P(z,t)$ is determined by a 
reaction-diffusion equation, eqn 4, with certain boundary conditions. 
The solution to this equation can be presented in the form
\begin{equation}
P(z, t) = \sum_{n=1}^{\infty} e^{\lambda_n t} \psi_n(z) \ ,
\label{eq:sum}
\end{equation}
where $\psi_n$ and $\lambda_n$ are the eigenfunctions and eigenvalues of the equation
\begin{equation}
(\mu(z) - m  - \lambda_n) \psi_n + D \frac{d \psi(z)}{d z^2} = 0 
\label{eq:boundprobl}
\end{equation}
with the same boundary conditions for the functions $\psi_n(z)$. 
The eigenvalue $\lambda^*$ with the largest real part, the so-called principal eigenvalue,
is of special interest for determining 
the dynamics of $P(z,t)$ because the term with this eigenvalue dominates the sum \ref{eq:sum} at large times (see e.g., Ryabov \& Blasius 2008). 
The sign of the real part of the principal eigenvalue determines if the population 
will grow, and so is able to persist
$\lambda^*>0$, or if it declines exponentially on the whole habitat, $\lambda^*<0$.
Note, that $\lambda^*$ depends on the spatial resource profiles and therefore may change in time with depletion of resources by a growing population.
If a species has attained equilibrium, it has zero population net growth and its principal eigenvalue equals zero, $\lambda^*=0$. 

The principal eigenvalue can be used for invasion analysis in a spatially extended system.
We can define a family of invaders which for small initial density in the presence of the resident species have zero net growth, and thus a zero principal eigenvalue $\lambda^*_{inv}=0$. 
In the space of the invader's critical resource values, this family defines the invasion threshold 
(in general, a hypersurface), separating successful and unsuccessful invaders. 
In our setting, and given that resident and invader have identical maximal growth rate, $\mu_{{\rm max}, 2} = \mu_{{\rm max}, 1}$, and mortality, $m_2=m_1$,
this corresponds to the condition 
$\Delta = c_{N, 1}^{-1}\,  \ln H_{N, 2}/H_{N, 1} + c_{I, 1}^{-1} \, \ln H_{I, 2}/H_{I, 1} = 0$ 
(see eqn~11).

Now consider an invader which has the same half saturation constants as the resident species, but different maximal growth rate, $\mu_{{\rm max}, 2}$, mortality, $m_2$, or dispersal ability, $D_2$.
It can be shown (Stakgold, 2000) that the eigenvalue spectrum,
and in particular the principal eigenvalue,
of this species shifts in the positive direction with an increase of the growth rate difference $\mu_{max, 2} - \mu_{max, 1}$,
and in the negative direction with an increase in the difference of destructive factors, such as dispersal rates $D_2-D_1$ or mortalities $m_2-m_1$. 
Thus, in general, the invader's principal eigenvalue will be positive or negative, even though the constant  remains $\Delta = 0$ (since the half saturation constants remain unchanged).
Assume without loss of generality that the invader has a positive total growth rate, $\lambda^*_{inv} > 0$. This means that it should be possible to invade the system with even higher resource requirements. 
Thus, there should be 
a `border-line' invader with the same values of 
$\mu_{{\rm max}, 2}$,  $m_2$, and $D_2$, but 
larger half saturation constants so that its principal eigenvalue equals zero, $\lambda^*_{inv}=0$. As the resource requirements of this second invader are higher, its $\Delta$ value is positive.
This value, which we denote by $\Delta^*_{12}$,
defines the difference in the favorable ranges (according to eqn~11) between this species and the resident. 
Thus all species with maximal growth rate $\mu_{\rm max, 2}$, dispersal rate $D_2$ and mortality $m_2$, 
and whose half-saturation constants satisfy the inequality
\begin{equation}
\frac{1}{c_{N, 1}}  \ln \frac{H_{N, 2}}{H_{N, 1}} + \frac{1}{c_{I, 1}}  \ln \frac{H_{I, 2}}{H_{I, 1}} < \Delta^*_{12} \ 
\label{eq:cond_gen} 
\end{equation}
can invade the system. 
As the boundary problem can not be solved in general, we cannot calculate the value of $\Delta^*_{12}$ but we can make some predictions about its sign. Similar to $\lambda^*_{inv}$, this value 
increases with $(\mu_{max, 2} - \mu_{max, 1})$, decreases with $m_2 - m_1$ and  $D_2 - D_1$, equals zero when all these differences vanish, and changes sign together with $\lambda^*_{inv}$.

Assuming Monod limitation of growth by two essential resources (eqn 3) we can rewrite eqn S12 
in terms of the critical resources values 
\begin{equation}
\frac{1}{c_{N, 1}}  \ln \frac{N^*_2}{N^*_1} + \frac{1}{c_{I, 1}}  \ln \frac{I^*_2}{I^*_1} < \Delta_{12} \ ,
\label{eq:cond_res_gen} 
\end{equation}
where 
\begin{equation}
\Delta_{12} = \Delta^*_{12} + \left(\frac{1}{c_{N, 1}} + \frac{1}{c_{I, 1}}  \right) 
   \ln  \frac{(\mu_{\max, 1} - m_1) m_2}{(\mu_{\max, 2} - m_2) m_1} \ .
\label{eq:cond_res_gen_delta} 
\end{equation}
Eqns. S13 and S14 define the invasion threshold for the invader, species 2, as a straight line 
with slope  $\gamma_1 = c_{N, 1}/c_{I, 1}$ 
in double logarithmic resource space.
The location of the invasion threshold depends in a complex way on the resource gradients and on the differences in maximal growth rate, $\mu_{\max, i}$, mortality, $m_{i}$, or dispersal rates, $D_{i}$. 
In the resource plane, the invasion threshold 
can be located above ($\Delta_{12} >0$) or below ($\Delta_{12} <0$) the 
critical resource values of the resident, $(N^*_1, I^*_1)$, see Fig~4a. 

\medskip

Consider a special case when 
the growth and mortality rates of species 2 are rescaled, 
$\mu_{\rm max, 2} = \beta \mu_{\rm max, 1}$ and $m_2 = \beta m_1$.
Then the second term in eqn \ref{eq:cond_res_gen_delta} vanishes, however 
$\Delta^*_{12}$ 
can be nonzero
because both $\mu_{max, 2} - \mu_{max, 1}$ and $m_2 - m_1$ are positive. 
Substituting $\mu_{\rm max, 2}$ and $m_2$ into  eqn~4
and dividing it by $\beta$, we obtain that in equilibrium the distribution of species 2 should satisfy the equation 
$$
\mu_{1}(z) P_2 - m_1 P_2 + \frac{D_1}{\beta} \frac{\partial^2 P_2}{\partial z^2} = 0 \ .
$$
Therefore these changes can be interpreted as a reduction of the diffusivity $D_2 =  D_1 / \beta$. 
Since $\Delta^*_{12}$ decreases with $(D_2-D_1)$ 
the shift of the invasibility threshold is positive ($\Delta^*_{12} > 0$) if $\beta > 1$ and negative  ($\Delta^*_{12} < 0$) if $\beta < 1$.

\section*{S4. Extension to systems with sinking or advection}

The dynamics of a population living in a unidirectional flow
(e.g. sinking, floating, etc.)
can be described by the reaction-diffusion-advection equation (Murray
2002, Ryabov \& Blasius 2008)
\begin{equation}
\frac{\partial P}{\partial t} = (\mu(z) - m) P -v \frac{\partial
P}{\partial z} + D \frac{\partial^2 P}{\partial z^2} \ ,
\label{eq:reactdiffadv}
\end{equation}
where $v$ is the flow or sinking velocity, which is positive when the flow is
directed towards larger $z$ values.
Substituting $P =   \tilde{P}  \exp \left( v z /2D\right)$ into eqn
\ref{eq:reactdiffadv}, we obtain
\begin{equation}
\frac{\partial \tilde{P}}{\partial t} = \mu \tilde{P} - \left( m +
\frac{v^2}{4D} \right) \tilde{P}  +
 D \frac{\partial^2 \tilde{P}}{\partial z^2} \ .
\label{eq:tilde_equ}
\end{equation}
Note that $\partial_t \tilde{P} $ has the same sign as $\partial_t P$,
so that both functions grow and decline simultaneously. 
Furthermore, they have the same boundary conditions
$\tilde{P}(0)=\tilde{P}(Z_B)=0$. 
Eqn.~\ref{eq:tilde_equ} describes a population $\tilde{P} (z,t)$ in a system without flow but with an effective
mortality
\begin{equation}
m' = m +  \frac{v^2}{4D} \  .
\label{eq:adv_mort}
\end{equation}
Therefore, the presence of an advective flow which washes out a population from a
favorable range can be interpreted as an effective increase of the
mortality by the value $v^2/4D$. 

This higher mortality entails an increase of resource
requirements. Assuming eqn 3 for the growth rate, we obtain new
critical resource values
\begin{equation}
I'^*_i =  H_{I, i}\frac{m'_i}{\mu_{max,i} - m'_i}\ , \qquad N'^*_i =
H_{N, i} \frac{m'_i}{\mu_{max,i} - m'_i} \ .
\label{eq:adv_mort}
\end{equation}
Since $m' > m$, the new critical resources values $I'^*$ and
$N'^*$ in the presence of
a positive flow are larger than in a system without flow.  This, in turn 
shifts the zero net growth isoclines and invasion thresholds towards higher resource values in the resource plane.

\newpage

\section*{S5. Model details and parameters used for figures}

{\bf Numerical scheme}
The model was integrated using a backward difference method, based on the finite volume scheme (Pham Thi et al. 2005). For the numerical solution we have discretized all variables on a grid which consisted of 400 points.
Diffusion terms were approximated by a second order central discretization scheme and 
integration was made via the trapezoidal rule. The resulting system of
ordinary differential equations was solved by the CVODE package
(http://www.netlib.org/ode).
Further we verified that the results remain unchanged if we double the number of points in the grid. 
See Ryabov et al. (2010) for further details on numerical procedures.

{\bf Boundary conditions}
As boundary conditions we assumed impenetrable borders at the surface and at the bottom for the phytoplankton biomass and an impenetrable surface and a constant concentration $N_B$ at the bottom for the nutrient.

{\bf Initial conditions}
We used a linear initial nutrient distribution changing from 0 at the surface to $N_B$ at the bottom.
For both phytoplankton species we assigned a uniform distribution of small initial density. 
We also investigated the influence of different initial conditions, however we did not find any new solutions beside the ones described.\\
To perform the invasion analysis, in the numerics the growth of the invader species was suppressed during the first 10000 simulation days, to make sure that an equilibrium distribution of the resident species was established. Then we simulated the system for a duration of further 40000 days to obtain the final competition outcome. 
To test for bistability,
this simulation was repeated by a second simulation in which the roles of invader and resident were exchanged.\\[1.5 ex]
{\setlength\parindent{0pt}
{\bf Parameter values} 
Default  species and model parameters are listed in Table \ref{tab:par}.\\[1ex]
{\bf Figure~1d.}
Calculation of the invasion threshold is
based on 7000 simulations of invasion by species with different half saturation constants $H_N$ and $H_I$. 
However residents and invaders have the same $\mu_{\max}, \ m$, and $D$, so that the invasion threshold is not shifted with respect to the resident $(N^*, I^*)$ values.

{\bf Figure~2.} 
Half-saturation constants of the invader, $H_I^{\rm inv} =8\
\mu$mol photons m$^{-2}$ and $H_N^{\rm inv} =$0.09 mmol nutrient
m$^{-3}$.

{\bf Figure~3b.}
The figure shows the results of more than 1000 simulations where parameter values of environmental conditions 
($I_{in}$, $K_{bg}$, $D$, $N_B$)
and values of consumption rates, $k_{i}$ and $\alpha_{i}$, 
were randomly chosen from a uniform distribution in a wide range (see Table~S1).
Specific parameters, which do not vary, correspond to species 1 and 2 in Table~\ref{tab:par}.
In particular, the half-saturation constants, maximal growth and mortality rates 
were fixed, so that the critical values, $(N_i^*, \ I_i^*$), and $\gamma_{cr}$ remain constant.
To determine $\gamma_1$ and $\gamma_2$ numerically we performed 
simulations for each species in monoculture.
Typically a change in environmental parameters leads to a simultaneous
increase or decrease of $\gamma_1$ and $\gamma_2$. As a result all data points group along one diagonal.\\
In the case of competitive dominance (where one species wins and the other is excluded) the ratio $\rho=\ln B_1/B_2$ practically approaches positive or negative infinity
after our standardized simulation period of 40000 days (see above). 
To visualize intermediate values of $\rho$ close to the coexistence region, the color scale was truncated at absolute values of $|\rho\,| \geq 7$, i.e.
log-ratios above 7 where set to $\rho=7$ and 
log-ratios below -7 where set to $\rho=-7$.\\
To judge bistability, simulations were repeated two times, with either
species 1 taken as resident and species 2 as invader (upward-pointing triangles)
or species 2 as resident and species 1 as invader (downward-pointing triangles).

{\bf Figure~5.} 
Invasion thresholds are plotted on the basis of invasion analysis for 7000 combinations of species 1 and test species 2.
Parameter values 
taken as in Table~\ref{tab:par}. 
However, for species 2 half saturation constants were varied to achieve the chosen values of critical resources, and
maximal growth rate and mortality 
taken as
(a)  $\mu_{{\rm
max}, 2} = 0.04$ h$^{-1}$ and $m_2=0.01$ h$^{-1}$ and
(b) 
$\mu_{{\rm max}, 2} = 0.08$ h$^{-1}$ and $m_2=0.02$ h$^{-1}$.

}

\subsection*{References}

{\setlength\parindent{0pt}

Arndt, S.,  Brumsack, H.-J. \& Wirtz, K.W. (2006).
Cretaceous black shales as active bioreactors: A biogeochemical model for the deep biosphere encountered during ODP Leg 207 (Demerara Rise).
{\em Geochimica et Cosmochimica Acta}, 70, 408-425.\\[1mm]
Beckmann, A. \& Hense, I. (2007). Beneath the surface: Characteristics of oceanic ecosystems under weak mixing conditions - A theoretical investigation. {\em Prog. Oceanogr.}, 75, 771-796.\\[1mm]
D'Hondt, S., Jorgensen, B.B., Miller, D.J., Batzke, A., Blake, R., Cragg, B.A. et al. (2004). Distributions of Microbial Activities in Deep Subseafloor Sediments. {\em Science}, 306, 2216-2221.\\[1mm]
Karl, D.M. \& Letelier, R.M. (2008). Nitrogen fixation-enhanced carbon sequestration in low nitrate, low chlorophyll seascapes. {\em Mar. Ecol. Prog. Ser.}, 364, 257-268.\\[1mm]
Karl, D. (2010). 
Oceanic ecosystem time-series programs. 
{\em Oceanography}, 23, 104.\\[1mm]
Klausmeier, C. \& Litchman, E. (2001). Algal games: The vertical distribution of phytoplankton in poorly mixed water columns. {\em Limnol. Oceanogr.}, 46, 1998-2007.\\[1mm]
Litchman, E., Klausmeier, C.A., Schofield, O.M. \& Falkowski, P.G. (2007). The role of functional traits and trade-offs in structuring phytoplankton communities: scaling from cellular to ecosystem level. {\em Ecol. Lett.}, 10, 1170-1181.\\[1mm]
Murray, J.D. (2002)
{\em Mathematical biology: an introduction.}
Springer Verlag.\\[1mm]
Pham Thi, N.N., Huisman, J.,  \& Sommeijer, B.P. (2005).
Simulation of three-dimensional phytoplankton dynamics: competition
 in light-limited environments.
{\em Journal of Computational and Applied Mathematics}, 174, 83-96.\\[1mm]
Ryabov, A.B. \& Blasius, B. (2008). 
Population growth and persistence in a heterogeneous environment: the role of diffusion and advection. 
{\em Math. Model. Nat. Phenom.}, 3, 42-86.\\[1mm]
Ryabov, A.B., Rudolf, L. \& Blasius, B. (2010).
Vertical distribution and composition of phytoplankton under the influence of an upper mixed layer.
{\em J. Theor. Biol.}, 263, 120-133.\\[1mm]
Shipboard Scientific Party (2003) 
Leg 201 Summary. 
In: {\em Proceedings of the Ocean Drilling Program, Initial Reports} 201, 1--81. 
College Station TX. \\[1mm]
Shipboard Scientific Party (2004)
Leg 207 summary.
In: {\em Proceedings of the Ocean Drilling Program, Initial Reports} 207,  1--89. 
College Station TX.\\[1mm]
%
Stakgold, I. (2000). {\em Boundary value problems of mathematical physics.} SIAM.\\[1mm]
Torres, M.E., Brumsack, H.J., Bohrmann, G. \& Emeis, K.C. (1996)
Barite fronts in continental margin sediments: A new look at barium remobilization in the zone of sulfate reduction and formation of heavy barites in diagenetic fronts.
{\em Chemical Geology}, 127, 125-139.\\[1mm]

}


\newpage

\begin{figure}[tb]
\begin{center}
\centerline{
\includegraphics[width=0.6\columnwidth]{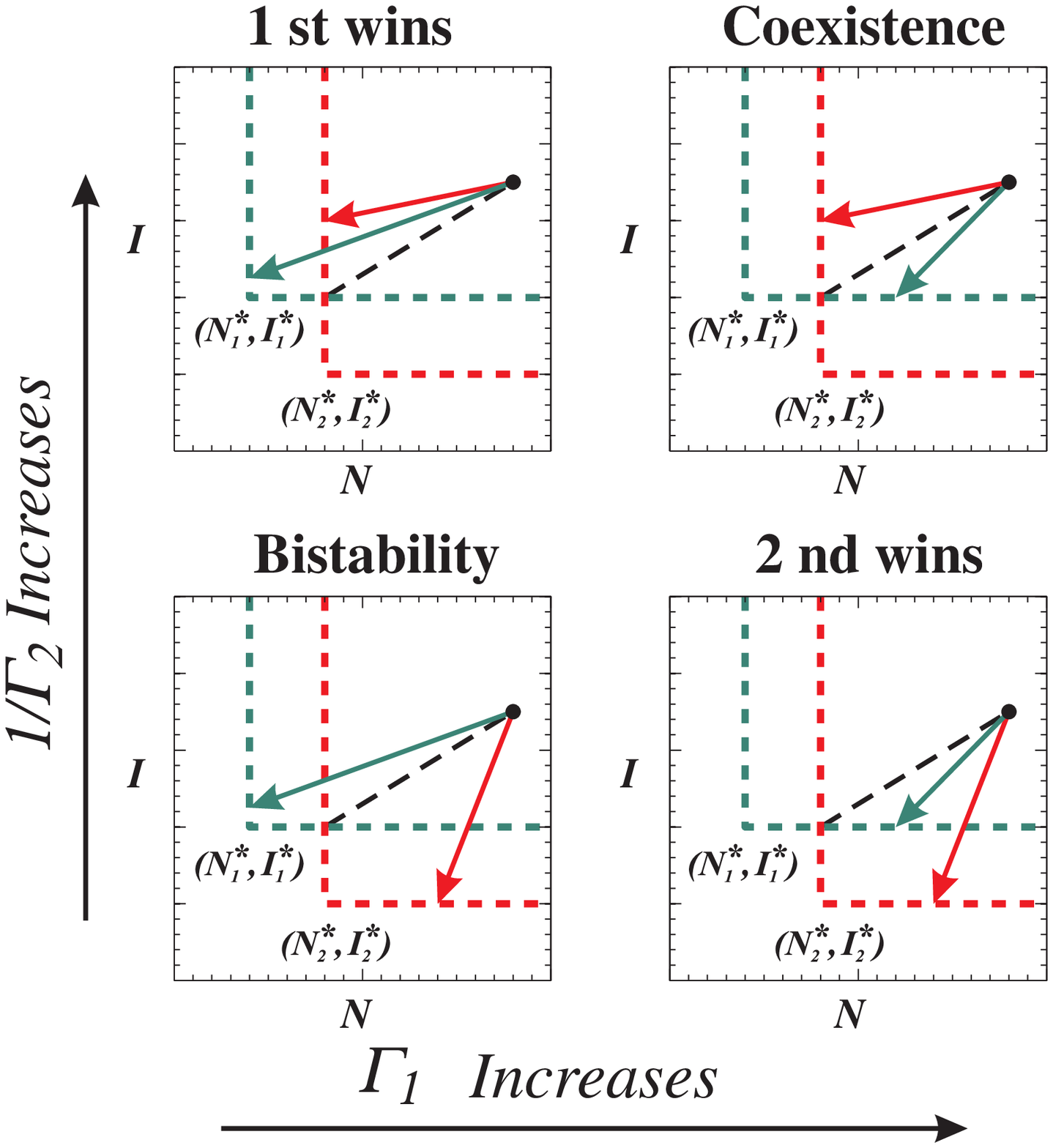}}
\caption{Two-species resource competition in a uniform system (compare to Fig.~3).
The four typical cases for the competition outcome are shown, depending on the mutual slopes of consumption vectors $\Gamma_i$
(values of critical resources and supply points remain fixed). 
For each case the zero-net-growth isoclines (ZNGI's, dashed lines) of species 1 (green) and species 2 (red) are plotted in the $(N,I)$-resource plane, together with the supply point (black circle) and the consumption vectors (green and red arrows). 
Each species in monoculture reduces resources to an equilibrium point, corresponding to the intersection of its consumption vector with its ZNGI. Invasion of the other species is possible if its critical resource values are located below this point (see Fig.~1b).
Define the critical slope $\Gamma_{cr}$ by the slope of a straight line from the supply point to the intersection of the two ZNGIs (indicated as black dashed line).
Species 1 can invade if the consumption vector of species 2 has a slope less than the critical slope ($1/\Gamma_2 > 1/\Gamma_{cr}$, top panel), i.e. species 2 has relatively stronger influence on its most limiting resource $N$ than on resource $I$.
In contrast, species 2 can invade if the slope of the consumption vector of species 1 is larger than the critical slope ($\Gamma_1 > \Gamma_{cr}$, right-hand panel), i.e. species 1 has a relatively stronger influence on its most limiting resource $I$.\\
}
\label{fig:4_Cases}
\end{center}
\end{figure}

\begin{figure*}[tb]
\begin{center}
\centerline{\includegraphics[width=1\textwidth]{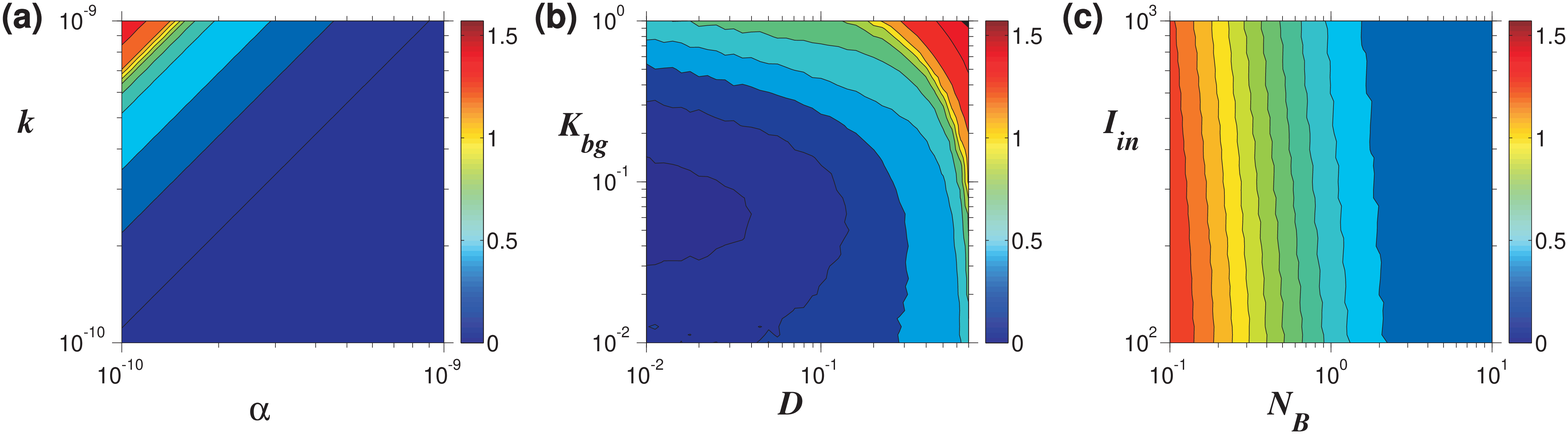}}
\caption{
Ratio of the logarithmic resource gradients $\gamma_1 = c_{I,1} / c_{N,1}$
in the phytoplankton model as a function of several species specific and environmental parameters. 
Colorcode indicates the value of $\arctan \gamma_1$
calculated  numerically for 
a monoculture of
species 1 in the phytoplankton model (see Table~\ref{tab:par} for parameters which do not vary).
Since $\gamma_i$ equals the slope of the invasion line in the resource plane,
this allows to project shifts in the species composition and to gain insight of how a change of one parameter can compensate for the influence of other parameters. 
For example, an increase of $\gamma$ corresponds to a clockwise rotation of the invasion threshold lines, 
thereby favoring the best nutrient competitor.\\ 
(a) Dependence of $\gamma$ on species traits.
Slope of the invasion threshold $\gamma$ grows with the ratio $k/\alpha$. This dependence is also evident from both limits in eqn \ref{eq:gamma_kbg}.
Together with  Fig.~3, this result confirms our suggestion that 
two-species coexistence is more probable if each species relatively less consumes its most limiting resource (see main text).\\ 
(b) Dependence of $\gamma$ on environmental conditions.
Slope of the invasion threshold $\gamma$ grows with turbulent diffusivity $D$ and background turbidity $K_{bg}$. 
This indicates that a turbid but weakly-mixed environment should result in the same species composition as stronger mixed but clean waters (while the total biomass will, of course, in general be different).\\ 
(c) Dependence of $\gamma$ on resource supply.
Increase of incident light intensity $I_{in}$ or of bottom nutrient concentration $N_B$ decreases $\gamma$ (thus leading to a counter-clockwise rotation of the invasion thresholds) and creates more favorable conditions for the best light competitor (compare to Fig.~\ref{fig:change}).
}
\label{fig:slope}
\end{center}
\end{figure*}

\begin{figure*}[tb]
\begin{center}
\centerline{\includegraphics[width=0.7\textwidth]{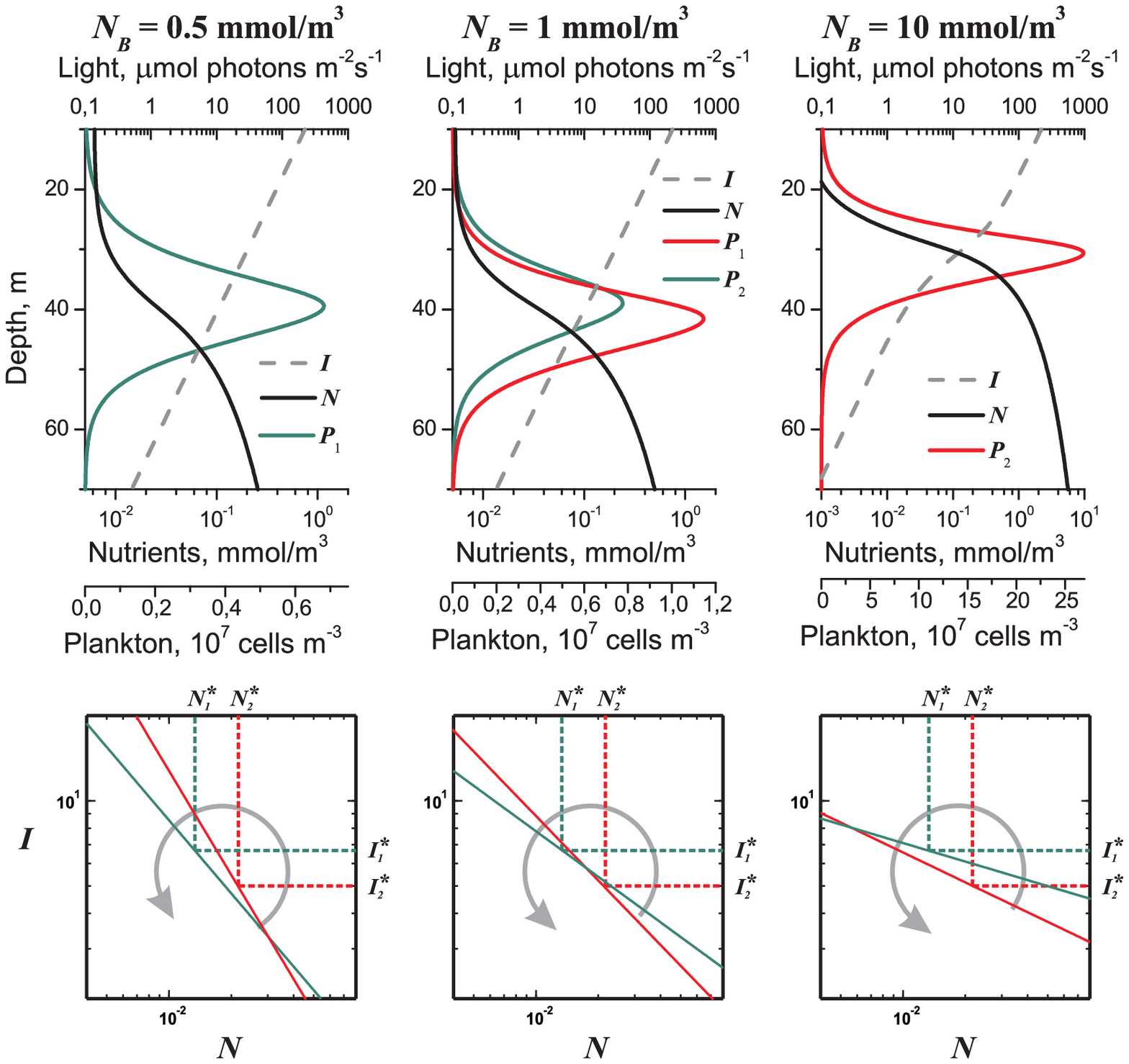}}
\caption{
Geometrical method to project the outcome of spatial resource competition in dependence of the ambient nutrient concentration, $N_B$.
Top panel: shift in the competition outcome between species 1 (green) and species 2 (red) in the phytoplankton model 
caused by an increase of the bottom nutrient concentration, $N_B$. Bottom panel illustrates this shift as the result of a counter-clockwise rotation of the invasion thresholds. The slopes, $\gamma_{1}$ and  $\gamma_{2}$, of the invasion thresholds in the bottom panel were calculated numerically for a monoculture of species 1 and 2. To obtain a relatively small difference between  $\gamma_{1}$  and  $\gamma_{2}$ 
we used $\alpha_1 = 8 \times 10^{-10}$  mmol nutrient cell$^{-1}$ and
$k_2 = 6 \times$10$^{-10}$ m$^2$ cell$^{-1}$.
}
\label{fig:change}
\end{center}
\end{figure*}

\begin{figure}[tbh]
\begin{center}
\centerline{\includegraphics[width=\textwidth]{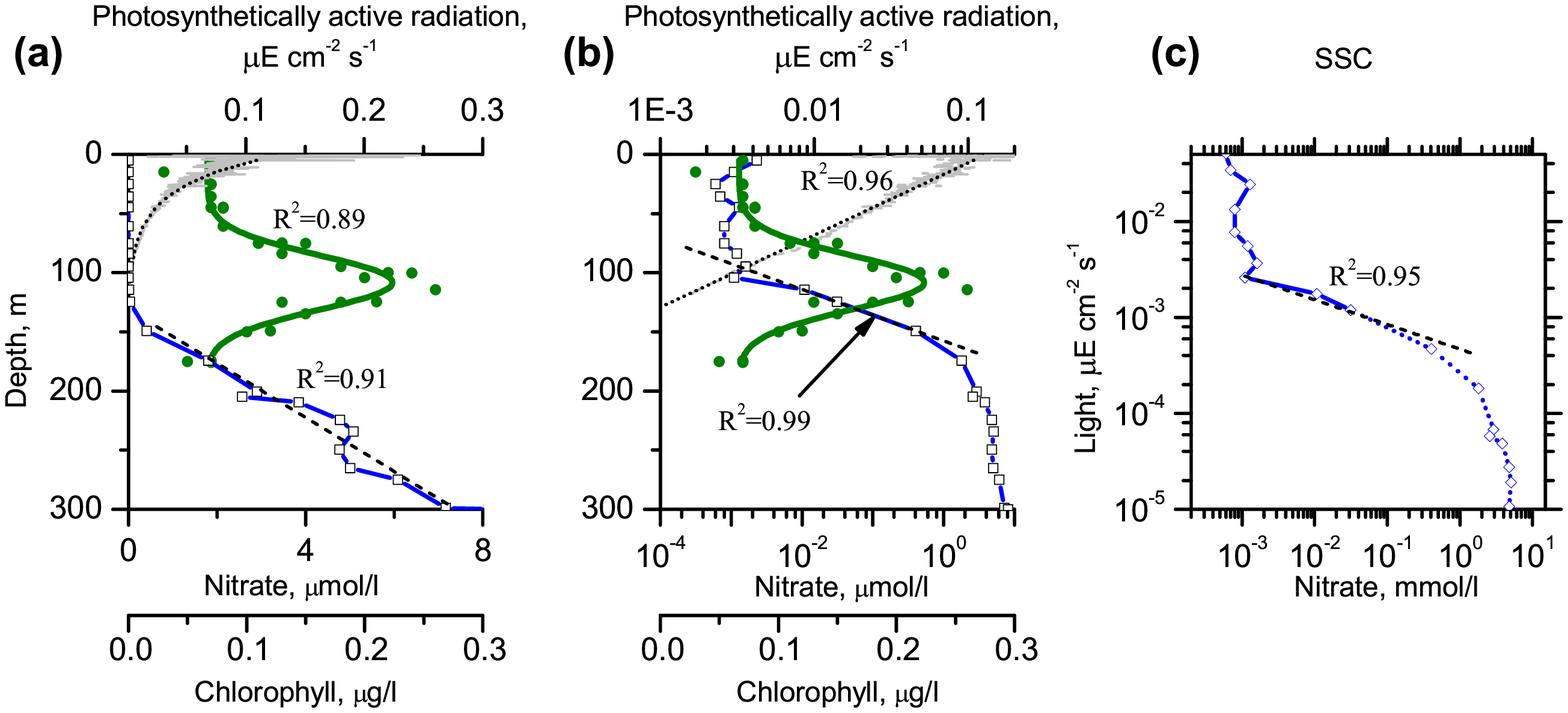}}
\caption{
Exponential resource distributions in a vertical water column.
Shown are the  profiles of photosynthetically active radiation (gray), nitrate (blue) and chlorophyll (green) measured during the HOT program (http://hahana.soest.hawaii.edu,  station ALOHA, cruise 114;
see also Karl 2010). Resource distributions plotted (a) on  linear and (b) on logarithmic scales. 
As it is commonly supposed, the light intensity can be approximated by an exponential distribution (black dotted lines and $R^2$ values in (a) and (b)).
In contrast, the nutrient profile exhibits two distinct regimes:
In the area below the production layer
the nutrient concentration can be well fit by a linear dependence ($R^2=0.91$, black dashed line in (a)); 
however, within the production layer
an exponential distribution gives a better description
($R^2=0.99$, black dashed line in (b)).
We argue that the regime of exponential decay within the production layer is more crucial for competition, 
because in this area growth can be nutrient limited, whereas
in the linear part of the nutrient profile growth is nutrient saturated.
Interestingly, the interval of exponential decay ranges over three orders of nitrate concentrations, making this area crucial for the competition among a wide range of phytoplankton species which can be limited by this nutrient (Litchman et al. 2007). 
Note, that similar exponential distributions in this interval of depths have been observed in 
mean nitrate concentrations between 1989-2006 (Karl \& Letelier 2008).
(c) System state curve (SSC), showing the distributions in the resource plane with logarithmic axes.
The solid blue line shows actual cruise data for the light intensity and the nutrient. Since light intensities are not reported below a depth of 140~m,
the system state curve was extrapolated, assuming that the light intensity follows the same exponential decay
up to 300 m (blue dotted line).
Dashed line shows a straight line fit (with $R^2$ value) within the production layer.
}
\label{fig:Hawaii}
\end{center}
\end{figure}

\begin{figure}[tbh]
\begin{center}
\centerline{\includegraphics[width=0.8\textwidth]{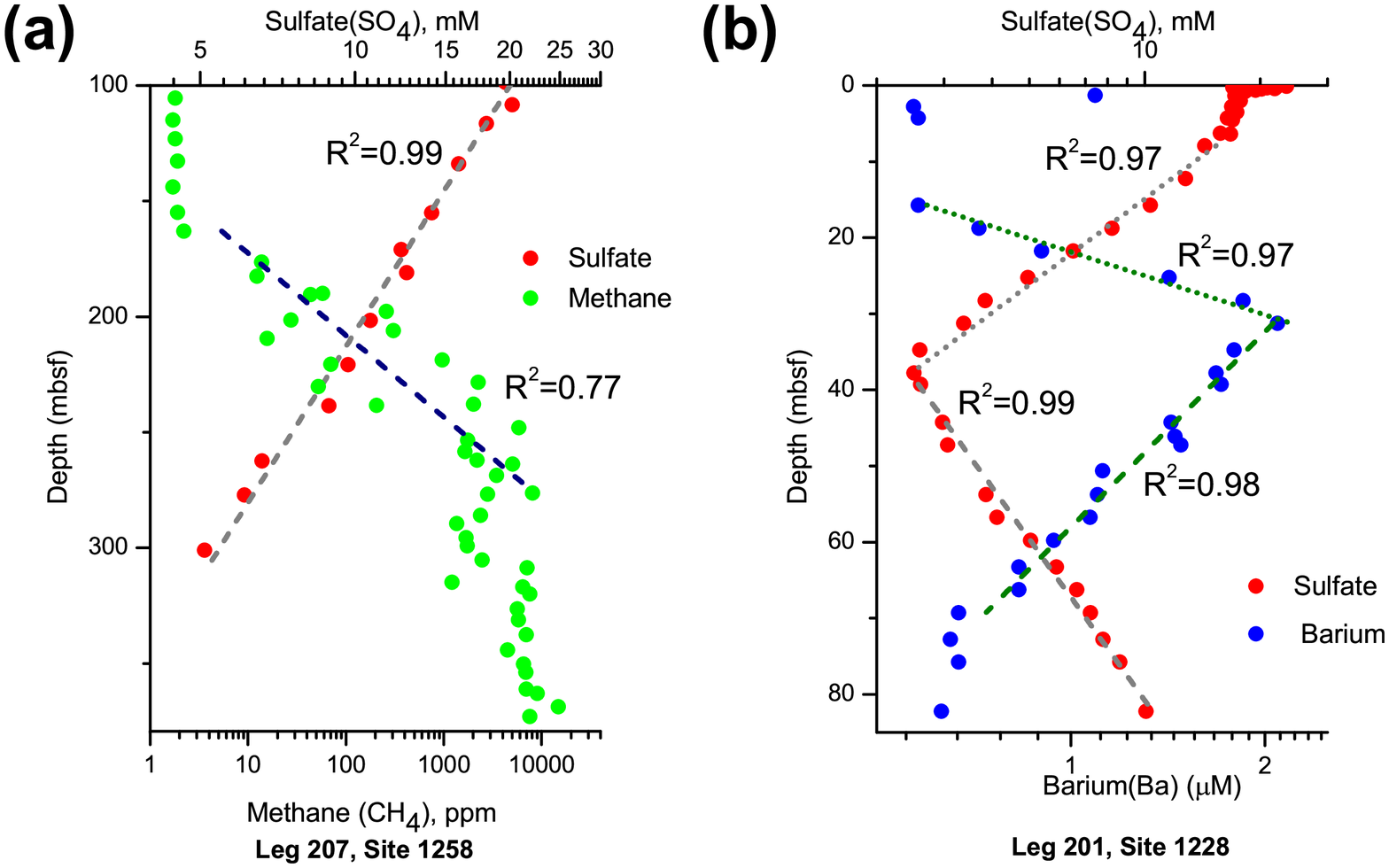}}
\caption{
Exponential distribution of resource concentrations in deep marine sediments.
Inverse concentration profiles (note the logarithmic scale) result from
biologically catalyzed reactions, which consume and release metabolites 
in a complex biogeochemical reaction network deep in the sediment column.\\
(a) Concentration profiles of sulfate (red) and methane (green) 
from ODP leg 207, site 1258, Demerara Rise, Equatorial Atlantic 
(Shipboard Scientific party, 2004;
Arndt et al. 2006)  
show a sulfate-methane transition zone
in a sediment depth of 150-300 m.
Substrates are supplied by a downward flux of sulfate from the sediment-water interface and
by the biogenic production of methane from deeply buried organic matter 
in black shale sequences at a depth below 400 m.
Upward diffusing methane and downward diffusing sulfate are depleted by
deep sedimentary microbial communities
in the process of anaerobic methane oxidation (AMO). 
Note that the exponential distribution of methane ranges over several orders of magnitude.\\
(b) Depth profiles of sulfate (red) and barium (blue) along the sediment column 
of ODP leg 201, site 1228, eastern Pacific ocean
(Shipboard Scientific party, 2004; D'Hondt et al, 2004), 
exhibiting two reversed zonation patterns.
The characteristic sulfate-barium transition zone
that extends from the water-sediment interface to greater depth 
(0 to 40 m) is mirrored by a second reversed succession 
that extends upward from the basement-sediment interface
in depth from 40 to 80 m.
Here, sulfate enters the sediment from two directions: from the overlying ocean 
and from an underlying basaltic aquifer.
In each transition zone the depletion of sulfate by microbial activity
promotes the remobilization of biogenic barium (Torres et al. 1996), giving rise
to high concentrations of dissolved barium in the pore fluids
beyond the zone of sulfate depletion.\\
Data are obtained during the ocean drilling program (ODP, http://iodp.tamu.edu/). 
Dashed  and dotted lines show straight line fits to the data on the logarithmic axes. $R^2$ values of the fit are indicated.
}
\label{fig:Sediments}
\end{center}
\end{figure}

\begin{figure}[tbh]
\begin{center}
\centerline{\includegraphics[width=0.65\textwidth]{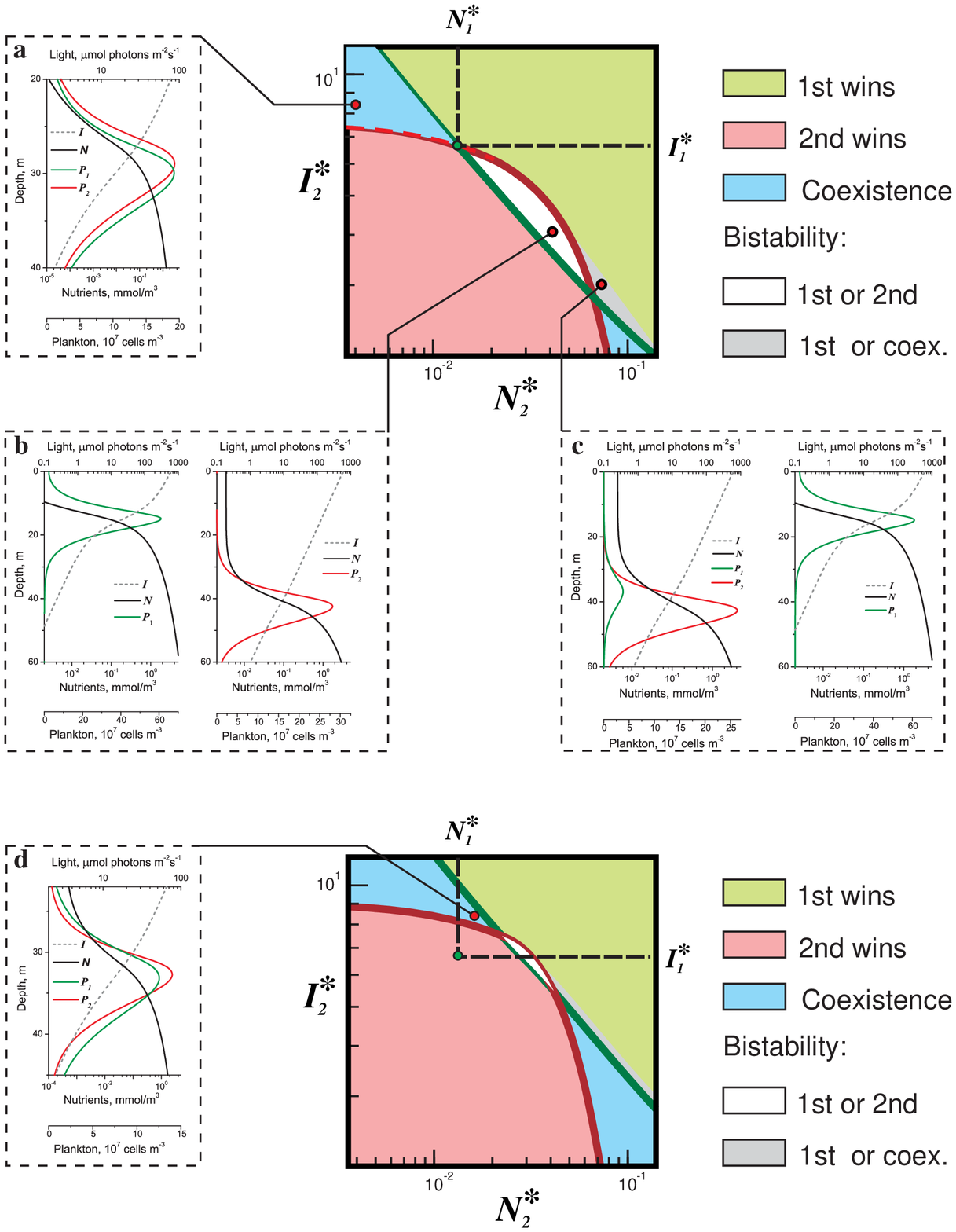}}
\caption{
Illustration of different competition outcomes in the phytoplankton model.
The outcome of two species competition is shown (indicated by color coding),
depending on the critical resource values of species 2 
(critical resources $(N_1^*,I_1^*)$ of species 1 fixed, green circle), 
when the maximal growth rates and mortalities are identical (top panel), or differ by the factor $\beta=2$ (bottom panel), similar to Fig.~5 from the main text.
Insets (a) - (d) show typical spatial profiles 
corresponding to the different competition regimes (critical resource values of species 2 indicated as red circle).
Plotted are the phytoplankton concentration of species 1 (green) and species 2 (red), and the distribution of light (black dashed line) and nutrient (black solid line) as a function of depth $z$.\\
(a) and (d):
Two fundamentally different coexistence mechanisms in a spatial system. 
(a) Coexistence due to a resource limitation trade-off,
mediated by niche segregation in resource requirements which becomes apparent as a spatial separation of density profiles.
(d) Coexistence due to a gleaner-opportunist trade-off.
The two species are not spatially separated,
but species 1 with smaller resource requirements (gleaner) can utilize a larger favorable range, whereas the high growth rate of the stronger resource limited species 2 (opportunist) allows it to survive on a smaller spatial range.\\
(b) and (c): 
Two kinds of bistability in the competition outcome.
(b) Alternative stable state when each species cannot grow in the presence
of its competitor. 
(c) Alternative stable states of either coexistence
(species 1 can invade the monoculture of species 2,
but does not reach a high abundance)
or a monoculture of species 1 (species 2
cannot establish in the presence of species 1).\\
Parameter values of species 1 see in Table~\ref{tab:par}.
Simulation parameters for  species 2:
(a) $H_{I, 2}=24.7 \ \mu$mol photons m$^{-2}$  s$^{-1}$,  $H_{N,2}=0.012$
mmol nutrient m$^{-3}$,
(b) $H_{I, 2}=12.2 \ \mu$mol photons m$^{-2}$  s$^{-1}$,  $H_{N,2}=0.123$
mmol nutrient m$^{-3}$,
(c) $H_{I, 2}=9 \ \mu$mol photons m$^{-2}$  s$^{-1}$,  $H_{N,2}=0.22$
mmol nutrient m$^{-3}$,
(d) $H_{I, 2}=24.7 \ \mu$mol photons m$^{-2}$  s$^{-1}$,  $H_{N,2}=0.044$
mmol nutrient m$^{-3}$.
}
\label{fig:Bistab}
\end{center}
\end{figure}

{
\linespread{1}

\begin{table*}[tbp]
\begin{center}
\caption{Default model parameters}
\begin{tabular}{llll}
\hline
\hline
Symbol & Interpretation  &  Units & Value \\ \hline
\multicolumn{4}{l}{\textbf{Independent variables}}\\
$t$ & Time & h &  \\
$z$ & Depth & m &  \\
\hline
\multicolumn{4}{l}{\textbf{Dependent variables}}\\
$P(z, t)$ & Population density & cells m$^{-3}$ &  \\
$I(z, t)$ & Light intensity & $\mu$mol photons m$^{-2}$ s$^{-1}$ &  \\
$N(z, t)$ & Nutrient concentration & mmol nutrient m$^{-3}$ &  \\
\hline
\multicolumn{4}{l}{\textbf{Parameters}}\\
$I_{\rm in}$ & Incident light intensity & $\mu$mol photons m$^{-2}$
s$^{-1}$ & 600 $(100 \dots 1000)^*$ \\[0.5ex]
$K_{\rm bg}$ & Background turbidity    & m$^{-1}$ & 0.1 $(0.01 \dots
1)^*$ \\ [0.5ex]
$Z_B$  & Depth of the water column     & m & 100 \\ [0.5ex]
$D$  & Vertical turbulent diffusivity& cm$^2$ s$^{-1}$ & 0.3 $(0.01
\dots 0.7)^*$ \\ [0.5ex]
$N_B$  & Nutrient concentration at $Z_B$ & mmol nutrient m$^{-3}$ & 10
$(0.1 \dots 10)^*$ \\[0.5ex]
\hline
\multicolumn{4}{l}{\textbf{Species 1}}\\
$H_I$ & half-saturation constant for light & $\mu$mol photons m$^{-2}$ \ s$^{-1}$  & 20 \\[0.5ex]
$H_N$ & half-saturation constant for nutrient & mmol nutrient m$^{-3}$ & 0.04 \\[0.5ex]
$k$ & light attenuation coefficient  &m$^2$ cell$^{-1}$ & 6$ \times$10$^{-10}$ 
$(1 \times$10$^{-10}$  \dots 1$ \times$10$^{-9})^*$ \\[0.5ex]
$\alpha$ & cell nutrient content & mmol nutrient cell$^{-1}$ & 2$\times$10$^{-10}$
$(1 \times$10$^{-10}$  \dots 1$ \times$10$^{-9})^*$ \\[0.5ex]
$\mu_{\rm max}$ & maximal growth rate & h$^{-1}$ & 0.04 \\[0.5ex]
$m$ & mortality rate & h$^{-1}$ & 0.01 \\[0.5ex]
\multicolumn{4}{l}{\textbf{Species 2}}\\
$H_I$ & half-saturation constant for light &$\mu$mol photons m$^{-2}$ \ s$^{-1}$ & 15 \\[0.5ex]
$H_N$ & half-saturation constant for nutrient & mmol nutrient m$^{-3}$ &  0.065 \\[0.5ex]
$k$ & light attenuation coefficient &m$^2$ cell$^{-1}$ & 1$ \times$10$^{-10}$ 
$(1 \times$10$^{-10}$  \dots 1$ \times$10$^{-9})^*$ \\[0.5ex]
$\alpha$ & cell nutrient content & mmol nutrient cell$^{-1}$ &  5$\times$10$^{-10}$ 
$(1 \times$10$^{-10}$  \dots 1$ \times$10$^{-9})^*$ \\[0.5ex]
$\mu_{\rm max}$ &   maximal growth rate & h$^{-1}$ & 0.04 \\[0.5ex]
$m$ & mortality rate & h$^{-1}$ & 0.01 \\[0.5ex]
\hline    
\multicolumn{4}{l}{$^*$ in 
Fig.~3b
and Fig.~\ref{fig:slope}.}\\
\multicolumn{4}{l}{Appendix S5 provides further details of simulation parameters specific to certain figures.}\\
\end{tabular}
\label{tab:par}
\end{center}
\end{table*}

}